\documentclass[acmsmall]{acmart}
\usepackage{enumitem}
\usepackage{makecell}
\usepackage{multirow}
\usepackage{fontawesome}
\usepackage{subcaption}
\usepackage{hyperref}
\usepackage{xcolor}
\usepackage{verbatim}
\usepackage{graphicx}
\usepackage{balance}
\usepackage{pifont}
\usepackage{rotating}
\usepackage{soul}
\usepackage{amsmath}
\usepackage[lined,boxed,linesnumbered,commentsnumbered,ruled]{algorithm2e}
\usepackage{color}
\usepackage{threeparttable}
\usepackage{adjustbox}
\usepackage{caption}
\usepackage{subcaption}
\usepackage[normalem]{ulem} 
\usepackage{colortbl}
\definecolor{mygray}{gray}{.9}

\newcommand{\Comment}[1]{}

\NewDocumentCommand{\framecolorbox}{oommm}
 {
  \IfValueTF{#1}
   {\IfValueTF{#2}
    {\fcolorbox{#3}{#4}{\makebox[#1][#2]{#5}}}
    {\fcolorbox{#3}{#4}{\makebox[#1]{#5}}}%
   }
   {\fcolorbox{#3}{#4}{#5}}%
 }

\AtBeginDocument{
  \providecommand\BibTeX{{
    \normalfont B\kern-0.5em{\scshape i\kern-0.25em b}\kern-0.8em\TeX}}}

\setcopyright{acmcopyright}

\usepackage[most]{tcolorbox}

\newcommand{\finding}[2]{
  \begin{tcolorbox}[
    colback=gray!10,   
    colframe=gray!70, 
    boxrule=0pt,       
    borderline west={2pt}{0pt}{gray}, 
    sharp corners,      
    enhanced
  ]
  \textbf{Finding #1:} #2
  \end{tcolorbox}
}

\begin{document}

\setcopyright{acmlicensed}
\acmJournal{TOSEM}
\acmYear{2025} \acmVolume{1} \acmNumber{1} \acmArticle{1} \acmMonth{12}\acmDOI{xxxxxx}

\title{On the Effectiveness of Training Data Optimization for LLM-based Code Generation: An Empirical Study}

\author{Shiqi Kuang}
\orcid{}
\affiliation{
  \institution{School of Computer Software, Tianjin University}
  \city{Tianjin}
  \country{China}
}
\email{kuangshiqi@tju.edu.cn}

\author{Zhao Tian}
\orcid{0000-0002-9316-7250}
\affiliation{
  \institution{School of Computer Software, Tianjin University}
  \city{Tianjin}
  \country{China}
}
\email{tianzhao@tju.edu.cn}

\author{Tao Xiao}
\orcid{}
\affiliation{
  \institution{Kyushu University}
  \city{Fukuoka}
  \country{Japan}
}
\email{xiao@ait.kyushu-u.ac.jp}

\author{Dong Wang}
\orcid{}
\affiliation{
  \institution{School of Computer Software, Tianjin University}
  \city{Tianjin}
  \country{China}
}
\email{dong_w@tju.edu.cn}

\author{Junjie Chen}
\authornote{Junjie Chen is the corresponding author.}
\orcid{0000-0003-3056-9962}
\affiliation{
  \institution{School of Computer Software, Tianjin University}
  \city{Tianjin}
  \country{China}
}
\email{junjiechen@tju.edu.cn}

\begin{abstract}
Large language models (LLMs) have achieved remarkable progress in code generation, largely driven by the availability of high-quality code datasets for effective training. 
To further improve data quality, numerous training data optimization techniques have been proposed; however, their overall effectiveness has not been systematically evaluated. 
To bridge this gap, we conduct the first large-scale empirical study, examining five widely-used training data optimization techniques and their pairwise combinations for LLM-based code generation across three benchmarks and four LLMs.
Our results show that data synthesis is the most effective technique for improving functional correctness and reducing code smells, although it performs relatively worse on code maintainability compared to data refactoring, cleaning, and selection. 
Regarding combinations, we find that most combinations do not further improve functional correctness but can effectively enhance code quality (code smells and maintainability). 
Among all combinations, data synthesis combined with data refactoring achieves the strongest overall performance.
Furthermore, our fine-grained analysis reinforces these findings and provides deeper insights into how individual techniques and their combinations influence code generation effectiveness. 
Overall, this work represents a first step toward a systematic understanding of training data optimization and combination strategies, offering practical guidance for future research and deployment in LLM-based code generation.
\end{abstract}



\include{coverletter}
\clearpage

\maketitle
\section{Introduction} 
\label{sec:intro}

Code generation aims to automatically produce source code that satisfies natural language descriptions. 
Effective code generation can substantially improve software development efficiency~\cite{pandey2024transforming, sherje2024enhancing, jiang2024survey}, enhance software quality~\cite{martinovic2024impact}, and reduce repetitive development efforts~\cite{uyanik2023developing}.
In recent years, code generation has witnessed rapid advances in both academia and industry~\cite{wang2025survey, chen2021evaluating, dong2023codep, li2022competition, shen2022incorporating}, particularly with the emergence of large language models (LLMs), which have demonstrated remarkable potential for understanding natural language and generating executable code~\cite{soliman2024leveraging, jiang2024survey}.
To further improve LLM-based code generation, prior studies have explored a variety of optimization techniques, including prompt engineering~\cite{wang2023code4struct, chen2021evaluating, li2025large, gao2024preference}, chain-of-thought reasoning~\cite{li2025structured, zhu2025uncertainty, zhang2022automatic}, requirement engineering~\cite{jin2024mare, mu2024clarifygpt, tian2025aligning, khot2022decomposed}, and multi-agent frameworks~\cite{islam2024mapcoder, huang2023agentcoder}. Alongside these inference- and interaction-level approaches, training data optimization has emerged as a fundamental and increasingly influential direction~\cite{kim2024datarecipe, li2025scar, yu2024makes, improta2025quality}.

Recent studies suggest that the strong performance of advanced code generation models is largely driven by the availability of large-scale, high-quality code data used during pretraining and fine-tuning~\cite{jain2023improving, li2023textbooks, zhou2023lima, cao2023instruction}. 
High-quality training data enables LLMs to better align natural language requirements with corresponding code implementations, thereby improving both functional correctness and static code quality. 
However, existing code datasets often suffer from issues such as distribution imbalance, inconsistent coding styles, and data noise~\cite{2024The}. 
These issues can substantially degrade training effectiveness and model performance, and have become one of the most critical challenges for LLM-based code generation. 
As a result, training data optimization has always been a key fundamental strategy for improving the model performance in code generation~\cite{tsai2024code, zhang2024codefort}.

To address these challenges, existing studies have proposed various training data optimization techniques, which can generally be categorized into five classes based on their objectives and optimization methods.
\textit{Data augmentation} aims to expand training datasets through semantic-preserving transformations~\cite{yu2022data, shi2023cocosoda, dong2025boosting, dong2024gencode, jain2021contrastive}, embedding-level composition~\cite{dong2023mixcode, li2022exploring}, and retrieval-based methods~\cite{chen2024data}.
\textit{Data cleaning} aims to remove redundant~\cite{singh2024brevity} or low-quality~\cite{yang2023decoding, wang2024your, improta2025quality, xue2025clean, nandani2023dacos, cao2024codecleaner} samples through dynamic execution or static checking~\cite{liang2023cupcleaner, improta2025quality}.
\textit{Data selection} aims to identify high-quality subsets based on criteria such as quality~\cite{li2025scar, dong2024gencode, jainllm, xie2023data}, diversity~\cite{wang2024your}, or complexity~\cite{lv2025data}.
\textit{Data synthesis} leverages the powerful distillation model to generate higher-quality code from problem descriptions~\cite{yu2024wavecoder, luowizardcoder, li2024instructcoder, naduaș2025synthetic} or test cases~\cite{shao2025case2code, ma2025unitcoder}, as well as to generate higher-quality problem descriptions from existing implementations~\cite{wei2024magicoder, wu2025inversecoder}.
\textit{Data refactoring} aims to improve code readability~\cite{kim2024datarecipe, chakraborty2022natgen, bandarupalli2025ai, cao2024codecleaner} and semantic richness~\cite{jainllm}, producing more consistent and maintainable training datasets. 
Each category offers a distinct perspective on improving training data quality and, consequently, model performance in code generation.
Despite their promise, these training data optimization techniques have not yet been systematically evaluated in a unified setting. 
Existing empirical studies and surveys typically focus on individual optimization techniques, such as data selection~\cite{albalak2024survey} or data augmentation~\cite{dong2025boosting, zhuo2023source}, without providing a comprehensive comparison across multiple techniques.
More importantly, the complementarity, applicability, and combined effects of different training data optimization techniques remain poorly understood. 
This lack of unified benchmarking and integrative analysis limits both the practical adoption of these techniques and a deeper understanding of their roles in LLM-based code generation.


To bridge this gap, we conduct the first large-scale, systematic empirical study that comprehensively compares the effectiveness of training data optimization techniques for LLM-based code generation. 
Specifically, we select representative and state-of-the-art techniques from five major categories and further investigate the synergistic effects of combining different techniques. 
Under a unified experimental framework, we evaluate these techniques on three widely-used benchmarks and two representative LLM families (Qwen and Llama) with varying model sizes, using a diverse set of metrics that capture functional correctness, code smells, and code maintainability.
To structure our investigation, we formulate the following three research questions:

\textbf{RQ1: How effective are different training data optimization techniques for LLM-based code generation?}
Almost all five training data optimization techniques improve functional correctness, with data synthesis delivering the largest gains, improving \textit{Pass@1} by 60.80\%, 74.06\%, and 12.20\%, and \textit{AvgPassRatio} by 29.00\%, 35.00\%, and 11.00\% on APPS, CodeContests, and MBPP, respectively.
In addition, all techniques reduce code smells, with data augmentation, data selection, and data synthesis achieving the most consistent improvements, yielding average relative Code Smell Score increases of 22.69\%, 25.70\%, and 39.92\%, respectively.
Moreover, data refactoring, data selection, and data cleaning improve code maintainability (Maintainability Index gains of 4.15\%, 8.12\%, and 2.73\%), whereas data synthesis and data augmentation increase code complexity, reducing Maintainability Index by 5.05\% and 4.19\%, respectively.

\textbf{RQ2: How effective are combinations of training data optimization techniques for LLM-based code generation?} 
Most combined techniques improve functional correctness, but gains are driven almost entirely by data synthesis, achieving 30.48\%–42.66\% \textit{Pass@1} and 15.35\%–18.92\% \textit{AvgPassRatio} improvements. 
No combination consistently outperforms data synthesis alone across all models and benchmarks, with the combination of data synthesis and data refactoring (Syn+Ref) being the closest.
Nearly all combinations substantially reduce code smells, yielding average relative improvements of 18.02\%–63.27\% in Code Smell Score. 
Among them, Syn+Ref demonstrates the strongest effect.
Combinations involving data refactoring, data selection, and data cleaning further improve code maintainability, with average relative gains of 5.76\%–9.51\% in the Maintainability Index. 
In contrast, combinations that include data synthesis or data augmentation exhibit limited or even negative effects on code maintainability.

\textbf{RQ3: How do code modification magnitude and technique complementarity affect the effectiveness of training data optimization?}
Our fine-grained correlation analysis shows that the effective code modification magnitude significantly influences functional correctness in both individual and combination settings, particularly for combinations involving data synthesis.
Moreover, the analysis indicates that combining techniques with different yet complementary optimization mechanisms tends to yield superior performance, as exemplified by the combination of data synthesis and data refactoring. 
These insights further reinforce the findings observed in RQ1 and RQ2.

Overall, our study reveals several valuable takeaways: 
(1) \textit{Training data optimization works, but more is not always better}. 
Nearly all optimization techniques improve LLM-based code generation, underscoring data quality as a primary performance driver. 
However, combining multiple techniques rarely produces additive gains in functional correctness, revealing a clear upper bound and challenging the assumption that increased optimization necessarily leads to better outcomes.
(2) \textit{Complementarity matters more than quantity}.
Combining highly similar optimization techniques offers limited benefits, especially for functional correctness. 
In contrast, combining techniques that operate on different yet complementary optimization mechanisms, such as data synthesis and data refactoring, yields stronger overall performance, highlighting the importance of diversity and complementarity in data optimization pipelines.
(3) \textit{Align data optimization choices with target code quality objectives}. 
No single technique optimizes functional correctness, code quality, and code maintainability simultaneously. 
Particularly, data synthesis is most effective for improving functional correctness and reducing code smells, whereas data refactoring, data cleaning, and data selection better enhance code maintainability. 
Practitioners should therefore tailor data optimization choices to their specific quality goals rather than adopting a one-size-fits-all strategy.


\textbf{Contributions.} 
To sum up, we make the following major contributions: 
(1) We conduct the first large-scale empirical study of training data optimization techniques for LLM-based code generation.
(2) We systematically evaluate the effectiveness of both individual techniques and their combinations across 48 experimental configurations (i.e., 3 benchmarks $\times$ 4 LLMs $\times$ 4 metrics), providing actionable insights for researchers and practitioners.
(3) We publicly release a comprehensive replication package to facilitate reproducibility and support future research, available at \textcolor{violet}{\url{https://github.com/Kuangshiqi/Data-optimization-techniques}}.

\section{background and related work}
\subsection{LLM-based Code Generation}
Code generation has become one of the core tasks at the intersection of artificial intelligence and software engineering~\cite{hou2024large, soliman2024leveraging, bistarelli2025usage, jiang2024survey}.
Its goal is to enable models to automatically generate executable programs that meet both syntactic and semantic requirements based on natural language descriptions and input–output examples.

In recent years, LLMs have achieved remarkable progress in code generation. 
For instance, Meta AI's CodeLlama series~\cite{roziere2023code}, through targeted fine-tuning and reinforcement learning on multilingual code~\cite{wang2024enhancing}, achieves competitive performance against proprietary models across public benchmarks and offers developers high-quality code completion and refactoring services. 
Google DeepMind's Gemini series~\cite{team2023gemini} demonstrates strong inference efficiency, long-context understanding, and multimodal reasoning ability, making it suitable for large-scale code generation and testing tasks. 
Similarly, DeepSeek-R1~\cite{guo2025deepseek} substantially enhances reasoning performance and achieves results comparable to advanced closed-source models on mathematical, logical, and coding benchmarks.

Beyond these advancements, researchers have explored multiple directions to improve the performance of LLM-based code generation.
A straightforward approach is prompt engineering and chain-of-thought (CoT), which guides the model's reasoning process. 
For example, REDCODER~\cite{parvez2021retrieval} provides high-quality generation patterns via retrieval-augmented generation (RAG), while SCOT~\cite{li2025structured} encourages the model to produce structured chain-of-thoughts to optimize code generation.
Another important direction is requirement engineering-driven code generation, which supplies models with structured and well-defined specifications to reduce the ambiguity of natural language instructions. 
For instance, Specine~\cite{tian2025aligning} and ClarifyGPT~\cite{mu2024clarifygpt} clarify ambiguous instructions by aligning specifications or answering related questions.
The third direction involves multi-agent code generation frameworks, which support iterative reasoning and tool usage to simulate human programming processes, such as MapCoder~\cite{islam2024mapcoder} and AgentCoder~\cite{huang2023agentcoder}.
Orthogonal to these inference-time methods, \textbf{training data optimization} plays a fundamental role in improving model performance, as high-quality, diverse, and representative training instances directly determine the model effectiveness and generalization. 
Techniques such as SCAR~\cite{li2025scar}, which selects stylistically consistent code, and SPAT~\cite{yu2022data}, which augments code via semantically consistent transformations, highlight the importance of data-centric improvements~\cite{yu2024makes}.


\subsection{Training Data Optimization in Code Generation}


With the rapid advancement of LLMs in code generation tasks, the quality of training data has become a critical determinant of model performance. 
Low-quality, inconsistent, or noisy data not only degrades the correctness and robustness of generated code but may also cause models to internalize poor programming patterns, ultimately hindering their effectiveness in real-world scenarios~\cite{2024The, jain2023improving, tsai2024code}.
Consequently, researchers have increasingly turned their attention to optimizing low-quality training datasets, systematically improving data consistency to enhance model effectiveness and generalization~\cite{luo2025survey}.
Existing data optimization techniques can be broadly categorized into five types: data augmentation, data cleaning, data selection, data synthesis, and data refactoring.
Below, we provide an overview of each technique.

\begin{itemize}
    \item \textbf{Data Augmentation} expands the training dataset by rewriting or transforming existing code snippets to increase diversity and improve model robustness. 
    For example, SPAT~\cite{yu2022data} introduces a collection of semantics-preserving transformation rules that efficiently and accurately generate diverse code variants. 
    MIXCODE~\cite{dong2023mixcode} performs embedding-level augmentation by mixing representations of different samples to create novel training instances. 
    ~\citet{song2024code} achieve augmentation by enriching code with additional comments, thereby providing more contextual information for model learning.

    \item \textbf{Data Cleaning} focuses on identifying and removing noisy, incorrect, or low-quality samples to improve dataset reliability. 
    For example, ~\citet{improta2025quality} filters out code snippets containing compilation errors or severe warnings. 
    SCIP~\cite{yang2023decoding} performs embedding-level cleaning by detecting anomalous latent representations. 
    ~\citet{singh2024brevity} enhances dataset quality by removing excessively long and noisy code snippets. 
    Overall, data cleaning substantially improves the purity of training data and ensures more stable and effective model learning.

    \item \textbf{Data Selection} ranks and filters samples based on predefined metrics, retaining only the most informative subset of the original training dataset. 
    For example, SCAR~\cite{li2025scar} scores samples by measuring stylistic consistency between code snippets and selects those with high style alignment. 
    XCoder~\cite{wang2024your} evaluates factors such as code complexity, test coverage, and data diversity to guide the selection process. 
    ~\citet{lv2025data} introduces data-complexity metrics to ensure that the selected subset maintains distributional consistency with the original dataset. 
    By prioritizing high-quality samples, data selection not only enhances model performance but also reduces training costs.

    \item \textbf{Data Synthesis} leverages a powerful distillation model to synthesize new data, transferring their high-level programming knowledge into the training corpus.
    For example, WaveCoder~\cite{yu2024wavecoder} employs large models to generate code solutions from natural language descriptions. 
    Case2Code~\cite{shao2025case2code} produces code directly from test cases, while MagicCoder~\cite{wei2024magicoder} generates natural language descriptions from source code. 
    Overall, data synthesis enriches datasets with novel and consistent samples, thereby improving data quality, diversity, and representativeness.

    \item \textbf{Data Refactoring} applies semantics-preserving structural transformations to improve code quality, readability, and consistency. 
    For example, DataRecipe~\cite{kim2024datarecipe} integrates six code-normalization tools to refactor raw code into standardized representations. 
    NatGen~\cite{chakraborty2022natgen} enhances code naturalness using six types of semantics-preserving transformations. 
    CODECLEANER~\cite{cao2024codecleaner} mitigates data pollution through code restructuring and variable renaming. 
    In summary, data refactoring targets code-level deficiencies through structured optimization, effectively improving code quality, uniformity, and maintainability.
\end{itemize}
\section{Study Design}
\label{sec:study_design}
\begin{figure}[t]
    \centering
    \makebox[\textwidth]{
      \includegraphics[width=.9\textwidth,height=.8\textheight,keepaspectratio]{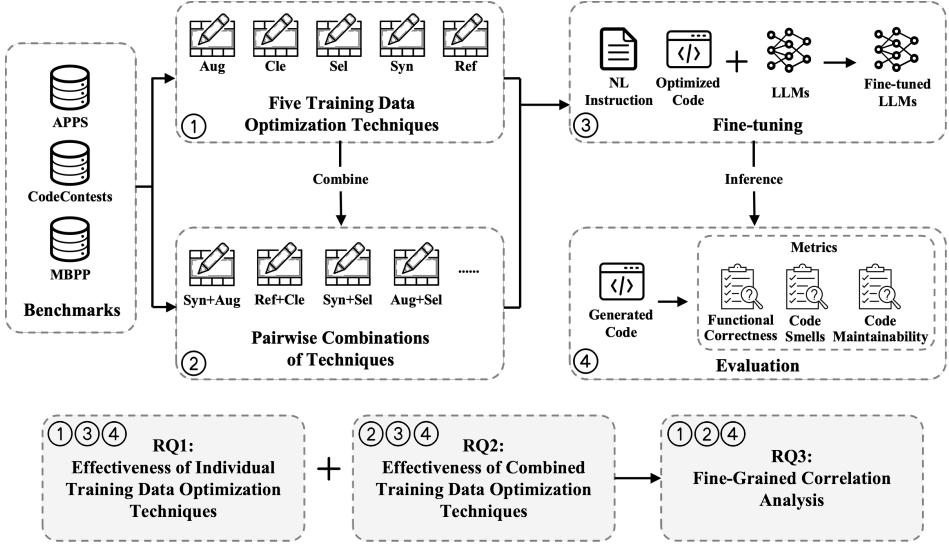}
    }
    \caption{The overview of our study design}
    \label{fig:overview}
\end{figure}

Figure~\ref{fig:overview} presents the overall design of our study. 
Using three widely-adopted benchmarks (i.e.,  APPS, CodeContests, and MBPP), we first investigate the effectiveness of individual training data optimization techniques on LLM-based code generation (\textbf{RQ1}).
Next, we evaluate the effectiveness of pairwise combinations of these techniques (\textbf{RQ2}).
Finally, by analyzing the complementarity and similarity of the datasets generated by each technique, we examine how their underlying characteristics relate to the observed improvements in model performance (\textbf{RQ3}). 

\subsection{Benchmarks}
\label{subsec:benchmarks}

To comprehensively evaluate the effectiveness of training data optimization techniques, we employ three widely-used benchmarks, i.e., APPS~\cite{hendrycks2measuring}, CodeContests~\cite{li2022competition}, and MBPP~\cite{austin2021program}. All three benchmarks have been extensively adopted in prior code generation studies~\cite{costello2025think, piterbargtraining, lei2025planning, islam2025codesim}.
Below, we provide an overview of each benchmark and describe the preprocessing steps applied to ensure consistency with the scope of our study.
Table~\ref{tab:benchmark} summarizes the statistical information about these studied benchmarks, including the training, validation, and test datasets.

\textbf{Benchmark I. APPS}. It comprises 10,000 programming problems written in Python collected from public competitive programming platforms (e.g., LeetCode and Kattis), covering different difficulty levels and consisting of 5,000 training and 5,000 test instances. 
Each instance consists of a natural-language problem description and, on average, 21 test cases. We exclude training instances that lack test cases, as such instances cannot be used to evaluate the functional correctness of the optimized code.
To balance evaluation cost and the generalizability of our conclusions, we further constructed new validation and test sets by randomly sampling 100 and 500 instances, respectively, from the original test set, following the difficulty distribution defined in the original dataset (Introduction: Interview: Competition = 1 : 3 : 1).

\noindent
\begin{table}[b]
    \centering
    \normalsize
    \tabcolsep=3.0mm
    \captionof{table}{Statistical summary of studied benchmarks}
    \label{tab:benchmark}
    \vspace{-0.8em}
    \begin{tabular}{lccc}
        \toprule
        \midrule
        \textbf{Benchmarks} & \textbf{\# Training} & \textbf{\# Validation} & \textbf{\# Test} \\
        \midrule
        \textbf{APPS}         & 4,337 & 100 & 500 \\
        \textbf{CodeContests} & 7,278 & 95  & 122  \\
        \textbf{MBPP}         & 374  & 90 & 500 \\
        \midrule
        \bottomrule
    \end{tabular}
    \vspace{0.3em}
\end{table}

\textbf{Benchmark II. CodeContests}. It is curated by DeepMind from the Codeforces programming platform, containing 13,328 training instances, 117 validation instances, and 165 test instances.
Each instance includes a natural language problem description and an average of 203.7 test cases, along with manually written reference solutions in multiple programming languages (i.e., Java, Python, and C++). 
Since our study focuses on Python, we excluded instances lacking Python3 implementations, yielding a filtered dataset with 7,278 training instances, 95 validation instances, and 122 test instances.

\textbf{Benchmark III. MBPP}. It is introduced by Google Research, containing 974 crowd-sourced Python programming problems, each with a natural language description and three test cases. 
Following the configuration of its original paper~\cite{austin2021program}, we divided the dataset into 374 training, 90 validation, and 500 test instances.


\subsection{Studied Training Data Optimization Techniques}
\label{subsec:training_data_optimization_methods}

In this study, following the prior work~\cite{yang2024large}, we selected one representative technique from each category in data augmentation, data cleaning, data selection, data synthesis, and data refactoring according to the following criteria:
\textit{(1) Open-Source or Reproducible.} 
To ensure the accuracy and reliability of our experimental results, we prioritized techniques that are open-source or can be replicated using the configurations reported in their original papers.
\textit{(2) Technique Coverage.} 
The selected techniques should target general-purpose code generation or be readily transferable to code generation tasks.
\textit{(3) Latest Studies.} 
To reflect the current state of the art, we favored recently proposed techniques, which generally demonstrate improved performance compared to earlier techniques.
The details of the selected techniques are presented below.

For the \textbf{Data Augmentation} technique, we selected the state-of-the-art SPAT~\cite{yu2022data} as the representative method. 
SPAT expands training data through 17 semantics-preserving program transformations originally designed for Java. 
Since our study focuses on Python, we adopt the 12 transformation rules that can be adapted to Python and apply each transformation to the original dataset. 
In particular, any transformed samples that fail to pass all test cases are discarded to further ensure code correctness.

For the \textbf{Data Cleaning} technique, we selected the latest technique by ~\citet{improta2025quality} as the representative approach. 
This method integrates Semgrep-based static analysis with dynamic testing to identify and remove low-quality training instances.
Semgrep~\cite{semgrep2025} is a widely-used industry-level static analysis tool capable of detecting common bugs, security vulnerabilities, and code quality issues. 
By combining static checks with execution-based validation, the technique filters out code that fails to compile or execute, as well as samples that exhibit severe warnings.
Following the same workflow, we applied both static and dynamic checks to all benchmarks and discarded any training instances containing errors or severe warnings.

For the \textbf{Data Selection} technique, we chose the most recent SCAR~\cite{li2025scar} as the representative method. 
SCAR analyzes the impact of code-style consistency and trains a ranking model to score training instances based on their stylistic similarity. 
Following SCAR's methodology, we apply its ranking model to all benchmarks and retain the top 20\% of instances with the highest code-style consistency scores.

For the \textbf{Data Synthesis} technique, we selected the up-to-date WaveCoder~\cite{yu2024wavecoder} as the representative method. 
WaveCoder adopts a generator–discriminator pipeline, in which an LLM generates code from natural language descriptions, and a discriminator filters the outputs based on correctness and diversity. 
To balance cost and performance, we use Gemini-2.0-flash as the generator, given its strong code-generation capability and significantly lower inference cost compared to larger frontier models. 
The discriminator validates generated samples using test cases, and any code that passes all tests within five attempts is added to the dataset; otherwise, the original sample is retained.

For the \textbf{Data Refactoring} technique, we selected the state-of-the-art DataRecipe~\cite{kim2024datarecipe} as the representative refactoring method. 
Following its original study, DataRecipe improves data quality by applying six automated refactoring tools, including formatters (Autopep8, Black, YAPF), code cleanup tools (Autoflake), docstring formatters (Docformatter), and quotation normalizers (Unify). 
Since Black and YAPF are more aggressive and may introduce semantic changes, we adopt a safer subset of four tools (Autopep8, Autoflake, Docformatter, and Unify) to refactor Python code while preserving code semantic correctness.

\subsection{Experimented Large Language Models}
\label{subsec:studied_models}

To assess the effectiveness of different training data optimization techniques, we selected four state-of-the-art open-source LLMs for code generation, covering both code-specific and general-purpose settings.
These models have been widely used in the recent literature~\cite{souza2025code, yan2025codeif, yan2025guiding, huang2024swiftcoder}, including: 

\begin{itemize}
    \item \textbf{Qwen2.5-Coder} is built upon the Qwen2.5 architecture and pretrained on a vast corpus of over 5.5 trillion tokens. 
    Qwen2.5-Coder series models have been evaluated on a wide range of code-related tasks, achieving state-of-the-art performance across more than 10 benchmarks, including code generation and repair. 
    Based on existing research~\cite{huang2024swiftcoder, yan2025codeif}, this study selects the Qwen2.5-Coder series models as the code-specialized LLM, aiming to leverage its deep understanding of code syntax, semantics, and programming conventions to achieve higher accuracy and professionalism.
    
    \item \textbf{Llama-3.2} is a series of LLMs based on the Transformer architecture, designed to improve performance across various language understanding tasks. 
    Based on existing research~\cite{souza2025code, yan2025codeif, yan2025guiding}, this study selected the Llama-3.2 series models as general-purpose LLMs, aiming to leverage their powerful general reasoning and natural language understanding capabilities to explore the ability of optimization strategies to solve complex programming problems.
\end{itemize}

To further examine how model size affects performance, we include two model sizes from each LLM family. 
To balance model effectiveness and computational resources, our evaluation covers four LLMs: Llama-3.2-1B-Instruct (LM-1B), Qwen2.5-Coder-1.5B-Instruct (QW-1.5B), Llama-3.2-3B-Instruct (LM-3B), and Qwen2.5-Coder-3B-Instruct (QW-3B).

\subsection{Model Training and Evaluation}
To systematically evaluate the effectiveness of the studied training data optimization techniques, we conducted large-scale experiments on three widely-used real-world code generation benchmarks and four representative LLMs. 
The overall experimental workflow consists of two main phases: \textbf{model training} and \textbf{model evaluation}.

In the \textbf{model training} phase, we first applied each training data optimization technique to the training split of each benchmark, thereby constructing a collection of optimized, higher-quality training datasets. 
Subsequently, for each optimized dataset, we performed supervised fine-tuning (SFT) on each LLM independently. 
During supervised fine-tuning, each training instance comprises a natural-language problem description as input and its corresponding reference code as the target output. 
Model parameters are optimized by minimizing the cross-entropy loss between the generated code and the reference code using gradient-based updates, progressively enhancing the model's code generation capability.
To ensure experimental consistency and fairness, we strictly controlled all training conditions across different models and datasets, including hyperparameter configurations, optimizer settings, and learning-rate scheduling strategies.
As a result, observed performance differences can be primarily attributed to the quality of the training data induced by different data optimization techniques, rather than confounding training factors. 
In addition, model performance was monitored on the validation split of each benchmark during training, and the checkpoint achieving the best validation performance was selected as the final fine-tuned model.

In the \textbf{model evaluation} phase, we evaluated all fine-tuned models on the original test split of each benchmark. 
By comparing test performance across different training data optimization techniques, we quantitatively analyzed their impact on code generation performance. 
Larger performance gains indicate more effective training data optimization strategies. 
Detailed experimental settings and metrics are further described in the sections corresponding to each research question.

\subsection{Evaluation Metrics}
\label{subsec:metrics}
To comprehensively assess code quality, we employed three categories of evaluation metrics (i.e., \textbf{functional correctness}, \textbf{code smells}, and \textbf{code maintainability}). 
These metrics are widely adopted in prior work~\cite{liu2025iterative, blyth2025static, li2025scar, improta2025quality, dong2025codescore, siahaan2025cyclomatic} and collectively provide a holistic view of the quality of code produced by LLM-based code generation techniques.

\smallskip
\noindent
\textbf{Metric I. Functional Correctness.}
Following existing work~\cite{kulal2019spoc, chen2021evaluating, hao2022aixbench, dong2025codescore, li2023skcoder}, we adopt \textit{Pass@K} and \textit{AvgPassRatio} to evaluate the effectiveness of training data optimization techniques in improving LLM-based code generation performance.

\textit{Pass@K} measures the functional correctness of the LLM-generated code based on the test execution. 
For each programming problem, the LLM generates K code instances. 
A problem is considered solved if at least one instance passes all associated test cases. 
Pass@K is defined as the percentage of solved problems among all problems.
Following the existing work~\cite{singh2024brevity, lv2024codeact, kim2024datarecipe, yang2023decoding, improta2025quality}, software developers typically consider and evaluate only a single code instance generated by the LLM-based code generation techniques, and thus we set K=1. 
In particular, Pass@1 is a strict and more challenging metric, as it requires the first generated solution to be fully correct. 
Larger Pass@K values indicate better code generation performance in terms of functional correctness. 

\textit{AvgPassRatio} measures the degree of partial correctness of generated code on associated test cases. 
Unlike Pass@K, which focuses solely on whether a generated code is completely correct on all test cases, AvgPassRatio measures the proportion of passed test cases for each generated code, averaged across all problems. 
Therefore, this metric is complementary to Pass@K, providing a more fine-grained view of the functional correctness. 
Larger AvgPassRatio values indicate better code generation performance in terms of functional correctness.

\smallskip
\noindent
\textbf{Metric II. Code Smells.}
Following prior work~\cite{liu2025iterative, blyth2025static, kim2024datarecipe}, we evaluated the \textit{Code Smell Score (CSS)} of generated code using Pylint~\cite{pylint2020}, a widely-adopted static analysis tool for Python.
Specifically, Pylint reports code smells issues across four categories:
(1) \textit{Error}: issues that may lead to program failure or obvious runtime errors;
(2) \textit{Warning}: potential bugs that may not cause immediate crashes but could lead to unexpected behavior;
(3) \textit{Refactor}: design or structural issues that suggest opportunities for better organization;
and (4) \textit{Convention}: style and readability issues, such as deviations from standard coding styles.
For each generated code, the CSS values are calculated as:
\begin{equation}
    CSS = \max\left(0, \; 10-10 \times \frac{w_{e} \times N_{e} + w_{w} \times N_{w} + w_{r} \times N_{r} + w_{c} \times N_{c}}{N_{s}}\right)
    \label{eq:css}
\end{equation}
where $N_{e}$, $N_{w}$, $N_{r}$, $N_{c}$, and $N_{s}$ denote the number of errors, warnings, refactor suggestions, convention issues, and the total number of statements in the code, respectively.
In particular, the corresponding weights $w_{e}$, $w_{w}$, $w_{r}$, and $w_{c}$ represent the severity of each issue type. 
Following previous work~\cite{pylint2020, pylintfeatures2021}, we set $w_{e}$=5.0 and assign a weight of 1.0 to all remaining categories.
Subsequently, we computed the CSS value for each generated solution and reported the average score across all programming problems. 
This metric provides a static-analysis–based perspective on code quality that complements functional correctness metrics derived from dynamic test execution.
Larger CSS values indicate better code generation performance in terms of code smells.

\smallskip
\noindent
\textbf{Metric III. Code Maintainability.}
Following existing work~\cite{oman1994construction,herivcko2023exploring,zaric2024software}, we adopt the Maintainability Index (MI) to further evaluate the maintainability of generated code.
MI is a widely-used quantitative indicator that provides a single-valued assessment of software maintainability, reflecting the ease of software to be maintained.
Specifically, the MI metric combines three established software engineering metrics:
(1) \textit{Halstead Volume (HV)}: captures lexical complexity and program length by quantifying the distinct operators and operands used in the code;
(2) \textit{Cyclomatic Complexity (CC)}: measures the number of independent control-flow paths, reflecting the complexity of logic branches and decision structures;
(3) \textit{Lines of Code (LOC)}: represents the program length, which is strongly associated with readability and maintainability.
For each generated code, the MI score is calculated as:
\begin{equation}
    MI = \max\left(0, \; 100 \times \frac{171 - w_{HV} \times \ln(\text{HV}) - w_{CC} \times \text{CC} - w_{LOC} \times \ln(\text{LOC})}{171}\right)
    \label{eq:mi}
\end{equation}
In particular, the coefficients $w_{HV}$, $w_{CC}$, and $w_{LOC}$ specify the contribution of each component, and following previous work~\cite{herivcko2023exploring}, we set $w_{HV}$=5.2, $w_{CC}$=0.23, and $w_{LOC}$=16.2.
We computed the MI score (ranging from 0 to 100) for each generated code and reported the average values across all programming problems. 
Larger MI score values indicate better code generation performance in terms of code maintainability.

\subsection{Implementation Details}
\label{subsec:implementation}
To replicate the studied data optimization techniques, we either adopted their publicly available implementations or re-implemented them strictly according to the descriptions provided in the original papers. 
Specifically, for data selection, data cleaning, and data refactoring, we used the reproduction packages or APIs released by the original techniques. 
For data augmentation and data synthesis, we manually reproduced the techniques following the original methodologies.

Throughout our experiments, we used identical hyperparameter settings for all models during both fine-tuning and inference. 
For fine-tuning, we set the learning rate to $2 \times 10^{-5}$ and employed the AdamW optimizer with a weight decay of 0.01. 
The maximum token length was limited to 1024, and the total batch size was set to 64 (i.e., batch size multiplied by gradient accumulation steps). 
Based on preliminary experiments, we fine-tuned all models for 10 epochs to ensure convergence while mitigating the risk of overfitting.

All open-source models (i.e., Llama-3.2-1B-Instruct, Qwen2.5-Coder-1.5B-Instruct, Llama-3.2-3B-Instruct, and Qwen2.5-Coder-3B-Instruct) were obtained from HuggingFace.\footnotemark[9] 
For the evaluated data synthesis technique (i.e., WaveCoder~\cite{yu2024wavecoder}), Gemini-2.0-flash was used as the base LLM and accessed via OpenRouter.\footnotemark[10] 
All experiments were conducted on a machine running Ubuntu 20.04, equipped with four NVIDIA H800 GPUs (80 GB each).

\begin{flushleft}
\footnotetext[9]{https://huggingface.co/}
\footnotetext[10]{https://openrouter.ai/}
\end{flushleft}

\section{Experimental RESULTS}
\label{sec:results}

\definecolor{myred}{RGB}{153,0,0}   
\definecolor{mygreen}{RGB}{0,102,0}  
\newcommand{\redtext}[1]{\textcolor{myred}{#1}}
\newcommand{\greentext}[1]{\textcolor{mygreen}{#1}}

\captionsetup[table]{skip=2pt}

\subsection{RQ1: Effectiveness of Individual Training Data Optimization Techniques}
\label{subsec:RQ1}

\smallskip
\noindent
\textbf{Approach.}
In this RQ, we evaluate the effectiveness of five training data optimization techniques (i.e., data augmentation, data cleaning, data selection, data synthesis, and data refactoring) individually across three representative benchmarks (APPS, CodeContests, and MBPP) using four code generation models (QW-1.5B, QW-3B, LM-1B, and LM-3B). 
This experimental design results in a total of 72 model configurations, i.e., 3 benchmarks $\times$ (5 techniques $+$ 1 original baseline) $\times$ 4 LLMs.
To assess effectiveness, we consider both functional correctness metrics (\textit{Pass@1} and \textit{AvgPassRatio}) and static code quality metrics (Code Smell Score and Maintainability Index), as described in Section~\ref{subsec:metrics}.


\begin{table*}[h]
\centering
\small
\caption{The effectiveness of five individual training data optimization techniques}
\label{tab:RQ1} 
\setlength{\tabcolsep}{2pt}
\begin{threeparttable}
\begin{adjustbox}{width=\textwidth}
\begin{tabular}{l cccc c cccc c cccc c}
\toprule
\midrule
\multirow{2}{*}{\textbf{Tech.}}
& \multicolumn{5}{c}{\textbf{APPS}}
& \multicolumn{5}{c}{\textbf{CodeContests}}
& \multicolumn{5}{c}{\textbf{MBPP}} \\
\cmidrule(lr){2-6} \cmidrule(lr){7-11} \cmidrule(lr){12-16} 
& LM-1B & QW-1.5B & LM-3B & QW-3B & \textbf{Avg.}
& LM-1B & QW-1.5B & LM-3B & QW-3B & \textbf{Avg.}
& LM-1B & QW-1.5B & LM-3B & QW-3B & \textbf{Avg.} \\
\midrule
\multicolumn{16}{l}{\framecolorbox[17.9cm][l]{gray!30}{gray!30}{\textbf{Pass@1 (Functional Correctness)}}} \\ \midrule
\textbf{Ori} & 3.80 & 8.60 & 9.40 & 13.40 & 8.80 & 4.10 & 3.28 & 10.66 & 7.38 & 6.36 & 23.40 & 44.60 & 37.80 & 50.00 & 38.95 \\
\textbf{Data Aug\tnote{*}} & 4.40\greentext{↑} & 9.80\greentext{↑} & 8.60\redtext{↓} & 13.80\greentext{↑} & 9.15\greentext{↑} & 4.10 & 6.56\greentext{↑} & 11.48\greentext{↑} & 5.74\redtext{↓} & 6.97\greentext{↑} & 22.00\redtext{↓} & 47.60\greentext{↑} & 36.40\redtext{↓} & 50.20\greentext{↑} & 39.05\greentext{↑} \\
\textbf{Data Cle} & 4.40\greentext{↑} & 10.00\greentext{↑} & 9.40 & 15.00\greentext{↑} & 9.70\greentext{↑} & 4.92\greentext{↑} & 4.92\greentext{↑} & 10.66 & 7.38 & 6.97\greentext{↑} & 25.20\greentext{↑} & 49.40\greentext{↑} & 39.40\greentext{↑} & 50.20\greentext{↑} & 41.05\greentext{↑} \\
\textbf{Data Sel} & 4.80\greentext{↑} & 11.40\greentext{↑} & 8.60\redtext{↓} & 13.00\redtext{↓} & 9.45\greentext{↑} & 5.74\greentext{↑} & 8.20\greentext{↑} & 13.11\greentext{↑} & 9.02\greentext{↑} & \underline{9.02}\greentext{↑} & 29.60\greentext{↑} & 43.60\redtext{↓} & 42.60\greentext{↑} & 50.00 & 41.45\greentext{↑} \\
\textbf{Data Syn} & 6.40\greentext{↑} & 14.40\greentext{↑} & 13.40\greentext{↑} & 22.40\greentext{↑} & \textbf{14.15}\greentext{↑} & 4.92\greentext{↑} & 11.48\greentext{↑} & 15.57\greentext{↑} & 12.30\greentext{↑} & \textbf{11.07}\greentext{↑} & 29.00\greentext{↑} & 50.60\greentext{↑} & 39.20\greentext{↑} & 56.00\greentext{↑} & \textbf{43.70}\greentext{↑} \\
\textbf{Data Ref} & 4.80\greentext{↑} & 11.20\greentext{↑} & 10.60\greentext{↑} & 13.40 & \underline{10.00}\greentext{↑} & 4.92\greentext{↑} & 4.10\greentext{↑} & 11.48\greentext{↑} & 7.38 & 6.97\greentext{↑} & 28.20\greentext{↑} & 47.20\greentext{↑} & 38.80\greentext{↑} & 53.60\greentext{↑} & \underline{41.95}\greentext{↑} \\
\midrule 
\multicolumn{16}{l}{\framecolorbox[17.9cm][l]{gray!30}{gray!30}{\textbf{AvgPassRatio (Functional Correctness)}}} \\ \midrule
\textbf{Ori} & 13.27 & 20.20 & 21.91 & 27.95 & 20.83 & 11.15 & 11.13 & 19.61 & 17.28 & 14.79 & 30.67 & 50.20 & 44.07 & 54.33 & 44.82 \\
\textbf{Data Aug} & 14.47\greentext{↑} & 23.05\greentext{↑} & 21.76\redtext{↓} & 28.23\greentext{↑} & 21.88\greentext{↑} & 11.85\greentext{↑} & 14.28\greentext{↑} & 18.29\redtext{↓} & 12.66\redtext{↓} & 14.27\redtext{↓} & 27.07\redtext{↓} & 53.00\greentext{↑} & 41.47\redtext{↓} & 54.33 & 43.97\redtext{↓} \\
\textbf{Data Cle} & 13.95\greentext{↑} & 22.39\greentext{↑} & 22.79\greentext{↑} & 28.79\greentext{↑} & 21.98\greentext{↑} & 12.66\greentext{↑} & 13.95\greentext{↑} & 21.69\greentext{↑} & 17.49\greentext{↑} & 16.45\greentext{↑} & 32.20\greentext{↑} & 54.33\greentext{↑} & 46.00\greentext{↑} & 55.93\greentext{↑} & 47.12\greentext{↑} \\
\textbf{Data Sel} & 13.47\greentext{↑} & 23.08\greentext{↑} & 19.40\redtext{↓} & 26.40\redtext{↓} & 20.59\redtext{↓} & 12.53\greentext{↑} & 17.59\greentext{↑} & 21.81\greentext{↑} & 16.12\redtext{↓} & \underline{17.01}\greentext{↑} & 37.20\greentext{↑} & 48.47\redtext{↓} & 48.60\greentext{↑} & 54.87\greentext{↑} & 47.29\greentext{↑} \\
\textbf{Data Syn} & 16.15\greentext{↑} & 27.74\greentext{↑} & 26.17\greentext{↑} & 37.47\greentext{↑} & \textbf{26.88}\greentext{↑} & 13.32\greentext{↑} & 20.07\greentext{↑} & 25.34\greentext{↑} & 21.08\greentext{↑} & \textbf{19.95}\greentext{↑} & 36.00\greentext{↑} & 56.20\greentext{↑} & 45.67\greentext{↑} & 61.13\greentext{↑} & \textbf{49.75}\greentext{↑} \\
\textbf{Data Ref} & 14.11\greentext{↑} & 23.88\greentext{↑} & 22.95\greentext{↑} & 27.16\redtext{↓} & \underline{22.03}\greentext{↑} & 14.24\greentext{↑} & 11.32\greentext{↑} & 21.76\greentext{↑} & 17.83\greentext{↑} & 16.29\greentext{↑} & 35.87\greentext{↑} & 52.20\greentext{↑} & 45.80\greentext{↑} & 57.87\greentext{↑} & \underline{47.94}\greentext{↑} \\
\midrule 
\multicolumn{16}{l}{\framecolorbox[17.9cm][l]{gray!30}{gray!30}{\textbf{CSS (Code Smells)}}} \\ \midrule
\textbf{Ori} & 5.45 & 5.15 & 5.69 & 5.42 & 5.43 & 7.00 & 5.44 & 6.73 & 5.94 & 6.28 & 2.36 & 0.81 & 1.29 & 2.17 & 1.66 \\
\textbf{Data Aug} & 5.40\redtext{↓} & 5.33\greentext{↑} & 6.22\greentext{↑} & 5.72\greentext{↑} & \underline{5.67}\greentext{↑} & 7.45\greentext{↑} & 7.32\greentext{↑} & 6.77\greentext{↑} & 6.85\greentext{↑} & \underline{7.10}\greentext{↑} & 2.45\greentext{↑} & 2.38\greentext{↑} & 3.15\greentext{↑} & 2.03\redtext{↓} & 2.50\greentext{↑} \\
\textbf{Data Cle} & 5.48\greentext{↑} & 5.76\greentext{↑} & 5.85\greentext{↑} & 5.53\greentext{↑} & 5.66\greentext{↑} & 7.19\greentext{↑} & 7.16\greentext{↑} & 6.87\greentext{↑} & 6.96\greentext{↑} & 7.05\greentext{↑} & 2.39\greentext{↑} & 1.61\greentext{↑} & 2.37\greentext{↑} & 2.19\greentext{↑} & 2.14\greentext{↑} \\
\textbf{Data Sel} & 5.27\redtext{↓} & 5.38\greentext{↑} & 5.68\redtext{↓} & 5.43\greentext{↑} & 5.44\greentext{↑} & 7.25\greentext{↑} & 7.51\greentext{↑} & 7.57\greentext{↑} & 6.81\greentext{↑} & \textbf{7.29}\greentext{↑} & 3.70\greentext{↑} & 1.83\greentext{↑} & 3.28\greentext{↑} & 1.88\redtext{↓} & \underline{2.67}\greentext{↑} \\
\textbf{Data Syn} & 5.60\greentext{↑} & 5.71\greentext{↑} & 6.20\greentext{↑} & 5.81\greentext{↑} & \textbf{5.83}\greentext{↑} & 6.58\redtext{↓} & 5.75\greentext{↑} & 6.38\redtext{↓} & 6.65\greentext{↑} & 6.34\greentext{↑} & 4.30\greentext{↑} & 3.48\greentext{↑} & 3.14\greentext{↑} & 3.10\greentext{↑} & \textbf{3.51}\greentext{↑} \\
\textbf{Data Ref} & 5.62\greentext{↑} & 5.68\greentext{↑} & 5.69 & 5.66\greentext{↑} & 5.66\greentext{↑} & 6.87\redtext{↓} & 7.00\greentext{↑} & 6.92\greentext{↑} & \textbf{7.27}\greentext{↑} & 7.02\greentext{↑} & 1.90\redtext{↓} & 1.85\greentext{↑} & 1.91\greentext{↑} & 1.92\redtext{↓} & 1.90\greentext{↑} \\
\midrule 
\multicolumn{16}{l}{\framecolorbox[17.9cm][l]{gray!30}{gray!30}{\textbf{MI (Code Maintainability)}}} \\ \midrule
\textbf{Ori} & 58.87 & 56.45 & 59.42 & 59.00 & 58.44 & 57.73 & 53.37 & 55.73 & 58.79 & 56.41 & 76.71 & 79.17 & 75.74 & 88.20 & 79.96 \\
\textbf{Data Aug} & 53.72\redtext{↓} & 55.36\redtext{↓} & 58.17\redtext{↓} & 57.74\redtext{↓} & 56.25\redtext{↓} & 54.77\redtext{↓} & 56.36\greentext{↑} & 53.60\redtext{↓} & 53.43\redtext{↓} & 54.54\redtext{↓} & 72.87\redtext{↓} & 77.41\redtext{↓} & 73.81\redtext{↓} & 78.10\redtext{↓} & 75.55\redtext{↓} \\
\textbf{Data Cle} & 58.85\redtext{↓} & 61.61\greentext{↑} & 62.93\greentext{↑} & 61.03\greentext{↑} & \underline{61.11}\greentext{↑} & 57.76\greentext{↑} & 57.09\greentext{↑} & 58.43\greentext{↑} & 57.20\redtext{↓} & 57.62\greentext{↑} & 77.47\greentext{↑} & 78.20\redtext{↓} & 81.03\greentext{↑} & 87.82\redtext{↓} & 81.13\greentext{↑} \\
\textbf{Data Sel} & 61.89\greentext{↑} & 64.33\greentext{↑} & 63.00\greentext{↑} & 63.96\greentext{↑} & \textbf{63.30}\greentext{↑} & 60.18\greentext{↑} & 56.36\greentext{↑} & 61.11\greentext{↑} & 55.18\redtext{↓} & \underline{58.21}\greentext{↑} & 89.53\greentext{↑} & 90.98\greentext{↑} & 88.19\greentext{↑} & 92.21\greentext{↑} & \textbf{90.23}\greentext{↑} \\
\textbf{Data Syn} & 54.67\redtext{↓} & 55.22\redtext{↓} & 55.01\redtext{↓} & 55.30\redtext{↓} & 55.05\redtext{↓} & 54.62\redtext{↓} & 49.84\redtext{↓} & 54.56\redtext{↓} & 52.29\redtext{↓} & 52.83\redtext{↓} & 77.32\greentext{↑} & 78.85\redtext{↓} & 74.69\redtext{↓} & 79.39\redtext{↓} & 77.56\redtext{↓} \\
\textbf{Data Ref} & 58.43\redtext{↓} & 57.95\greentext{↑} & 59.74\greentext{↑} & 58.45\redtext{↓} & 58.64\greentext{↑} & 58.82\greentext{↑} & 57.44\greentext{↑} & 56.23\greentext{↑} & 61.23\greentext{↑} & \textbf{58.43}\greentext{↑} & 84.17\greentext{↑} & 88.85\greentext{↑} & 83.86\greentext{↑} & 90.27\greentext{↑} & \underline{86.79}\greentext{↑} \\
\midrule
\bottomrule
\end{tabular}
\end{adjustbox}
\begin{tablenotes}
    \scriptsize
    \item[*] \parbox{0.75\linewidth}{
    Data Aug, Data Cle, Data Sel, Data Syn, and Data Ref are short for Data Augmentation, 
    Data Cleaning, Data Selection, Data Synthesis, and Data Refactoring, respectively.
    }
\end{tablenotes}
\end{threeparttable}
\end{table*}

\smallskip
\noindent
\textbf{Results.}
Table~\ref{tab:RQ1} presents the performance of five training data optimization techniques across twelve experimental settings (3 benchmarks $\times$ 4 LLMs). 
The ``Avg.'' column shows the average performance over four models, with bold indicating the best-performing technique and underlined indicating the second-best. ``\greentext{↑}'' and ``\redtext{↓}'' indicate results higher or lower than those fine-tuned on the original dataset, respectively. 

\textbf{Functional Correctness. }
The results show that nearly all training data optimization techniques improve functional correctness, with optimized training sets generally enhancing the models' ability to generate correct solutions. 
Among them, data synthesis yields the largest and most consistent gains, achieving average relative improvements of 60.80\%, 74.06\%, and 12.20\% in terms of \textit{Pass@1}, and 29.04\%, 34.89\%, and 11.00\% in terms of \textit{AvgPassRatio} on APPS, CodeContests, and MBPP, respectively. 
These improvements can be attributed to the use of training data synthesized by more advanced closed-source LLMs (e.g., Gemini-2.0-flash), which produce supervision signals that are easier for these smaller open-source models to learn. 
This observation is consistent with findings from prior work on data distillation and knowledge transfer~\cite{beyer2022knowledge, zhou2022large}.
In addition, data refactoring achieves the second-best performance on APPS and MBPP, with relative improvements of 13.64\% and 7.70\% in terms of \textit{Pass@1}, and 5.76\% and 6.96\% in terms of \textit{AvgPassRatio}, respectively. 
On CodeContests, data selection attains the second-best performance, yielding relative improvements of 41.82\% in terms of \textit{Pass@1} and 15.01\% in terms of \textit{AvgPassRatio}.
Other techniques also exhibit positive but relatively modest effects. 
For instance, data augmentation relatively improves \textit{Pass@1} by only 3.98\%, 9.59\%, and 0.26\% on APPS, CodeContests, and MBPP, respectively. 

\finding{I}{Nearly all data optimization techniques improve functional correctness in code generation tasks, with data synthesis achieving the largest and most consistent gains, yielding average relative improvements ranging from 12.20\% to 74.06\% in terms of \textit{Pass@1} and from 11.00\% to 34.89\% in terms of \textit{AvgPassRatio} across the three studied benchmarks.}

\textbf{Code Smells.}
The results show that almost all data optimization techniques tend to reduce code smells to some extent. 
Among them, data augmentation, data selection, and data synthesis exhibit the most consistent and pronounced effects. 
Specifically, data augmentation improves Code Smell Score (CSS) by an average of 4.42\%, 13.06\%, and 50.60\% across the three benchmarks. 
Data selection yields average improvements of 0.18\%, 16.08\%, and 60.84\%, while data synthesis achieves average gains of 7.37\%, 0.96\%, and 111.45\%, respectively.

These improvements can be attributed to the distinct characteristics of our studied techniques.
Firstly, data augmentation introduces code with diverse structures and styles, thereby expanding and balancing the distribution of structural and stylistic patterns in the training data. 
This reduces the likelihood that the model produces code smells due to unfamiliarity with a limited number of structures or styles. 
Secondly, data selection prioritizes higher-quality and stylistically consistent samples, which directly lowers the proportion of code smells in the training data and provides the model with cleaner and more stable learning signals. 
Thirdly, data synthesis leverages more advanced LLM to generate code with accurate, standardized solutions and consistent, well-structured coding styles; this high level of coding discipline effectively mitigates code smells, allowing synthesized data to serve as high-quality training samples.

In contrast, data refactoring primarily targets refactor- and convention-related categories in CSS, while paying less attention to error- and warning-related issues. 
Additionally, data cleaning mainly addresses error- and warning-related categories, and lacks systematic inspection and cleaning of refactor- and convention-related categories. 
As a result, their overall CSS improvements are less pronounced due to the lack of comprehensive optimization across all categories.
Overall, although the effectiveness of individual techniques varies across benchmarks and models, all techniques consistently reduce code smells, demonstrating that training data optimization is effective in improving static code quality.

\finding{II}{Almost all techniques achieve a reduction in code smells, with data augmentation, data selection, and data synthesis emerging as the most effective and consistently reliable, by average relative improvements of 22.69\%, 23.70\% and 39.92\%, respectively.
}

\textbf{Code Maintainability. }
The results show that techniques designed to improve code quality, including data refactoring, data cleaning, and data selection, effectively enhance code maintainability and reduce code complexity. 
Specifically, data refactoring relatively increases the Maintainability Index (MI) by an average of 0.34\%, 3.58\%, and 8.54\% across the three benchmarks. 
Data cleaning yields average MI improvements of 4.57\%, 2.15\%, and 1.46\%, while data selection achieves average gains of 8.32\%, 3.19\%, and 12.84\%, respectively.
In contrast, techniques based on incremental modification or complete rewriting, such as data augmentation and data synthesis, tend to increase code complexity. 
Data augmentation reduces the MI by an average of 3.75\%, 3.32\%, and 5.52\% across the three benchmarks, while data synthesis leads to average MI decreases of 5.80\%, 6.35\%, and 3.00\%, respectively.

This difference can be attributed to the inherent characteristics of the respective techniques. 
Specifically, data refactoring, data cleaning, and data selection improve code maintainability and reduce complexity by standardizing code structure, removing redundancy and noise, and prioritizing higher-quality samples, respectively. 
In contrast, data augmentation and data synthesis do not explicitly consider code complexity or code length during their optimization processes, which makes code maintainability less well guaranteed.

We further measured the average code length of the generated code produced by models fine-tuned with the five techniques across three benchmarks.
The results show that the code snippets generated based on data synthesis and data augmentation are the longest, with average lengths of 24.37 and 23.08 lines, respectively, whereas code snippets generated by the other techniques are more concise, with an average length of 19.16 lines. 
As indicated by Equation~\ref{eq:mi}, the MI value is negatively correlated with the lines of code (LOC). 
Consequently, the increase in code length leads to lower MI values for the data synthesis and data augmentation techniques, whereas the remaining three techniques exhibit improvements in MI.
In addition, since Halstead Volume (HV) and Cyclomatic Complexity (CC) are closely coupled with syntactic correctness and cannot be reliably isolated within the composite MI metric, we only report LOC values as a supplementary reference metric. 
Nevertheless, LOC alone is sufficient to substantiate our observations.

\finding{III}{Data refactoring, data cleaning, and data selection techniques improve code maintainability by increasing code quality, with average relative improvements ranging from 2.73\% to 8.12\%, while data augmentation and data synthesis generally decrease code maintainability, ranging from 4.19\% to 5.05\%.}

\begin{table*}[h]
\centering
\footnotesize
\caption{Model sensitivity to the individual training data optimization}
\label{tab:model_comparison}
\setlength{\tabcolsep}{4pt}
\begin{adjustbox}{width=\textwidth}
\begin{tabular}{l cc c cc c cc c cc} 
\toprule
\midrule
\multirow{2}{*}{\textbf{Technique}} 
& \multicolumn{2}{c}{\textbf{Pass@1}} 
& \multicolumn{2}{c}{\textbf{AvgPassRatio}} 
& \multicolumn{2}{c}{\textbf{CSS}} 
& \multicolumn{2}{c}{\textbf{MI}} \\
\cmidrule(lr){2-3} \cmidrule(lr){4-5} \cmidrule(lr){6-7} \cmidrule(lr){8-9}
 & QW Avg. & LM Avg. & QW Avg. & LM Avg. & QW Avg. & LM Avg. & QW Avg. & LM Avg. \\
\midrule
\textbf{Data Aug} & \textbf{+1.15} & -0.36 & -0.20 & \textbf{-0.96} & \textbf{+1.96} & +0.49 & \textbf{-3.32} & -2.88 \\
\textbf{Data Cle} & \textbf{+2.89} & +0.80 & \textbf{+3.61} & +1.44 & \textbf{+1.92} & +0.27 & \textbf{+5.67} & +2.05 \\
\textbf{Data Sel} & \textbf{+6.31} & +2.55 & \textbf{+5.16} & +2.06 & \textbf{+2.15} & +0.71 & \textbf{+10.32} & +6.62 \\
\textbf{Data Syn} & \textbf{+15.13} & +3.22 & \textbf{+16.06} & +3.66 & \textbf{+1.62} & +0.61 & \textbf{-7.55} & -2.22 \\
\textbf{Data Ref} & \textbf{+2.68} & +1.61 & \textbf{+2.97} & +2.34 & \textbf{+1.86} & +0.07 & \textbf{+6.33} & +2.84 \\
\midrule
\bottomrule
\end{tabular}
\end{adjustbox}
\end{table*}

\textbf{Model Sensitivity to Individual Training Data Optimization}. 
Furthermore, we examine the sensitivity of two model families to training data optimization techniques: the code-specialized Qwen2.5-Coder series and the general-purpose Llama-3.2 series. 
Table~\ref{tab:model_comparison} summarizes the average performance changes of the two model families under different data optimization techniques, where bold values indicate the model family exhibiting larger absolute performance increases or decreases.

The results show that Qwen2.5-Coder models are consistently more sensitive to both data optimization and data degradation operations. 
Specifically, in all improvement scenarios and in two out of three degradation scenarios, the magnitude of performance change for the Qwen2.5-Coder series is larger than that observed for the Llama-3.2 models. 
This behavior could be attributed to the stronger reliance of Qwen2.5-Coder models on code-specific patterns and semantic structures during training, which makes them more responsive to variations in data quality and distribution. 
As a result, these models tend to exhibit more pronounced performance fluctuations when the training data is either optimized or degraded.

\finding{IV}{Results suggest that code-specialized LLMs likely benefit more from high-quality data optimization, but are also more vulnerable to data degradation, highlighting the critical role of data curation when training or fine-tuning code-specialized models.}

\subsection{RQ2: Effectiveness of Combined Training Data Optimization Techniques}
\label{subsec:RQ2}

\begin{table*}[h]
\centering
\small
\caption{The effectiveness of ten pairwise combinations of various training data optimization techniques}
\label{tab:RQ2}
\setlength{\tabcolsep}{2pt}
\begin{threeparttable}
\begin{adjustbox}{width=\textwidth}
\begin{tabular}{l cccc c cccc c cccc c} 
\toprule
\midrule
\multirow{2}{*}{\textbf{Technique}}
& \multicolumn{5}{c}{\textbf{APPS}} 
& \multicolumn{5}{c}{\textbf{CodeContests}} 
& \multicolumn{5}{c}{\textbf{MBPP}} \\
\cmidrule(lr){2-6} \cmidrule(lr){7-11} \cmidrule(lr){12-16} 
 & LM-1B & QW-1.5B & LM-3B & QW-3B & \textbf{Avg.} 
 & LM-1B & QW-1.5B & LM-3B & QW-3B & \textbf{Avg.} 
 & LM-1B & QW-1.5B & LM-3B & QW-3B & \textbf{Avg.} \\
\midrule
\multicolumn{16}{l}{\framecolorbox[17.9cm][l]{gray!30}{gray!30}{\textbf{Pass@1 (Functional Correctness)}}} \\ \midrule
\textbf{Ori} & 3.80 & 8.60 & 9.40 & 13.40 & 8.80 & 4.10 & 3.28 & 10.66 & 7.38 & 6.36 & 23.40 & 44.60 & 37.80 & 50.00 & 38.95 \\
\textbf{Syn+Aug\tnote{*}}   & 6.80\greentext{↑} & 13.60\greentext{↑} & 12.00\greentext{↑} & 20.80\greentext{↑} & 13.30\greentext{↑} & 2.46\redtext{↓} & 13.11\greentext{↑} & 15.57\greentext{↑} & 11.48\greentext{↑} & \textbf{10.66}\greentext{↑} & 25.60\greentext{↑} & 49.00\greentext{↑} & 40.40\greentext{↑} & 55.20\greentext{↑} & 42.55\greentext{↑} \\
\textbf{Syn+Cle}   & 4.00\greentext{↑} & 14.80\greentext{↑} & 13.60\greentext{↑} & 21.00\greentext{↑} & \underline{13.35}\greentext{↑} & 3.28\redtext{↓} & 10.66\greentext{↑} & 11.48\greentext{↑} & 9.02\greentext{↑} & 8.61\greentext{↑} & 30.40\greentext{↑} & 51.00\greentext{↑} & 37.80 & 55.60\greentext{↑} & \underline{43.70}\greentext{↑} \\
\textbf{Syn+Sel}   & 6.40\greentext{↑} & 13.00\greentext{↑} & 13.20\greentext{↑} & 20.60\greentext{↑} & 13.30\greentext{↑} & 4.92\greentext{↑} & 6.56\greentext{↑} & 7.38\redtext{↓} & 14.75\greentext{↑} & 8.40\greentext{↑} & 29.60\greentext{↑} & 46.40\greentext{↑} & 42.20\greentext{↑} & 50.40\greentext{↑} & 42.15\greentext{↑} \\
\textbf{Syn+Ref}   & 8.40\greentext{↑} & 14.60\greentext{↑} & 13.20\greentext{↑} & 21.00\greentext{↑} & \textbf{14.30}\greentext{↑} & 3.28\redtext{↓} & 7.38\greentext{↑} & 15.57\greentext{↑} & 8.20\greentext{↑} & 8.61\greentext{↑} & 29.20\greentext{↑} & 49.20\greentext{↑} & 40.60\greentext{↑} & 56.40\greentext{↑} & \textbf{43.85}\greentext{↑} \\
\textbf{Ref+Aug}   & 4.40\greentext{↑} & 11.00\greentext{↑} & 10.00\greentext{↑} & 13.60\greentext{↑} & 9.75\greentext{↑} & 3.28\redtext{↓} & 6.56\greentext{↑} & 9.84\redtext{↓} & 8.20\greentext{↑} & 6.97\greentext{↑} & 23.20\redtext{↓} & 44.40\redtext{↓} & 35.00\redtext{↓} & 49.40\redtext{↓} & 38.00\redtext{↓} \\
\textbf{Ref+Cle}   & 4.20\greentext{↑} & 10.00\greentext{↑} & 9.20\redtext{↓} & 13.60\greentext{↑} & 9.25\greentext{↑} & 7.38\greentext{↑} & 7.38\greentext{↑} & 12.30\greentext{↑} & 7.38 & 8.61\greentext{↑} & 25.20\greentext{↑} & 48.80\greentext{↑} & 36.40\redtext{↓} & 50.40\greentext{↑} & 40.20\greentext{↑} \\
\textbf{Ref+Sel}   & 3.40\redtext{↓} & 10.20\greentext{↑} & 8.40\redtext{↓} & 13.60\greentext{↑} & 8.90\greentext{↑} & 5.74\greentext{↑} & 7.38\greentext{↑} & 11.48\greentext{↑} & 12.30\greentext{↑} & \underline{9.23}\greentext{↑} & 29.60\greentext{↑} & 43.00\redtext{↓} & 42.00\greentext{↑} & 50.20\greentext{↑} & 41.20\greentext{↑} \\
\textbf{Aug+Cle}   & 4.40\greentext{↑} & 10.00\greentext{↑} & 8.20\redtext{↓} & 13.00\redtext{↓} & 8.90\greentext{↑} & 5.74\greentext{↑} & 4.10\greentext{↑} & 13.93\greentext{↑} & 5.74\redtext{↓} & 7.38\greentext{↑} & 23.40 & 45.40\greentext{↑} & 33.20\redtext{↓} & 50.20\greentext{↑} & 38.05\redtext{↓} \\
\textbf{Aug+Sel}   & 3.60\redtext{↓} & 10.40\greentext{↑} & 10.00\greentext{↑} & 13.60\greentext{↑} & 9.40\greentext{↑} & 2.46\redtext{↓} & 2.46\redtext{↓} & 7.38\redtext{↓} & 7.38 & 4.92\redtext{↓} & 31.60\greentext{↑} & 41.80\redtext{↓} & 43.20\greentext{↑} & 49.20\redtext{↓} & 41.45\greentext{↑} \\
\textbf{Cle+Sel}   & 3.80 & 9.00\greentext{↑} & 8.60\redtext{↓} & 12.20\redtext{↓} & 8.40\redtext{↓} & 7.38\greentext{↑} & 9.02\greentext{↑} & 8.20\redtext{↓} & 5.74\redtext{↓} & 7.59\greentext{↑} & 29.60\greentext{↑} & 43.00\redtext{↓} & 39.00\greentext{↑} & 50.00 & 40.40\greentext{↑} \\
\midrule
\multicolumn{16}{l}{\framecolorbox[17.9cm][l]{gray!30}{gray!30}{\textbf{AvgPassRatio (Functional Correctness)}}} \\ \midrule
\textbf{Ori} & 13.27 & 20.20 & 21.91 & 27.95 & 20.83 & 11.15 & 11.13 & 19.61 & 17.28 & 14.79 & 30.67 & 50.20 & 44.07 & 54.33 & 44.82 \\
\textbf{Syn+Aug} & 16.16\greentext{↑} & 27.91\greentext{↑} & 25.72\greentext{↑} & 34.89\greentext{↑} & 26.17\greentext{↑} & 7.01\redtext{↓} & 20.10\greentext{↑} & 23.65\greentext{↑} & 21.39\greentext{↑} & \textbf{18.04}\greentext{↑} & 32.73\greentext{↑} & 55.47\greentext{↑} & 46.47\greentext{↑} & 61.00\greentext{↑} & 48.92\greentext{↑} \\
\textbf{Syn+Cle} & 12.51\redtext{↓} & 27.95\greentext{↑} & 27.21\greentext{↑} & 34.82\greentext{↑} & 25.62\greentext{↑} & 10.26\redtext{↓} & 18.52\greentext{↑} & 22.56\greentext{↑} & 17.81\greentext{↑} & 17.29\greentext{↑} & 37.20\greentext{↑} & 56.20\greentext{↑} & 44.87\greentext{↑} & 60.87\greentext{↑} & \textbf{49.79}\greentext{↑} \\
\textbf{Syn+Sel} & 16.58\greentext{↑} & 28.12\greentext{↑} & 27.31\greentext{↑} & 34.43\greentext{↑} & \underline{26.61}\greentext{↑} & 13.49\greentext{↑} & 11.43\greentext{↑} & 17.32\redtext{↓} & 23.49\greentext{↑} & 16.43\greentext{↑} & 37.20\greentext{↑} & 51.53\greentext{↑} & 48.13\greentext{↑} & 55.33\greentext{↑} & 48.05\greentext{↑} \\
\textbf{Syn+Ref} & 18.81\greentext{↑} & 28.05\greentext{↑} & 26.09\greentext{↑} & 34.44\greentext{↑} & \textbf{26.85}\greentext{↑} & 11.28\greentext{↑} & 14.34\greentext{↑} & 24.29\greentext{↑} & 17.80\greentext{↑} & 16.93\greentext{↑} & 35.47\greentext{↑} & 54.40\greentext{↑} & 47.13\greentext{↑} & 62.13\greentext{↑} & \underline{49.78}\greentext{↑} \\
\textbf{Ref+Aug} & 13.53\greentext{↑} & 23.95\greentext{↑} & 21.71\redtext{↓} & 26.10\redtext{↓} & 21.32\greentext{↑} & 9.24\redtext{↓} & 14.20\greentext{↑} & 18.31\redtext{↓} & 13.19\redtext{↓} & 13.74\redtext{↓} & 30.60\redtext{↓} & 50.13\redtext{↓} & 41.47\redtext{↓} & 54.53\greentext{↑} & 44.18\redtext{↓} \\
\textbf{Ref+Cle} & 13.82\greentext{↑} & 22.24\greentext{↑} & 22.63\greentext{↑} & 27.29\redtext{↓} & 21.50\greentext{↑} & 15.12\greentext{↑} & 16.37\greentext{↑} & 20.21\greentext{↑} & 17.65\greentext{↑} & 17.34\greentext{↑} & 33.33\greentext{↑} & 54.53\greentext{↑} & 43.40\redtext{↓} & 55.53\greentext{↑} & 46.70\greentext{↑} \\
\textbf{Ref+Sel} & 11.69\redtext{↓} & 22.67\greentext{↑} & 21.29\redtext{↓} & 27.47\redtext{↓} & 20.78\redtext{↓} & 13.27\greentext{↑} & 15.52\greentext{↑} & 21.05\greentext{↑} & 21.01\greentext{↑} & \underline{17.71}\greentext{↑} & 37.20\greentext{↑} & 48.00\redtext{↓} & 47.33\greentext{↑} & 54.93\greentext{↑} & 46.87\greentext{↑} \\
\textbf{Aug+Cle} & 14.25\greentext{↑} & 21.91\greentext{↑} & 21.06\redtext{↓} & 27.72\redtext{↓} & 21.24\greentext{↑} & 14.07\greentext{↑} & 13.82\greentext{↑} & 22.83\greentext{↑} & 14.37\redtext{↓} & 16.27\greentext{↑} & 29.53\redtext{↓} & 50.06\redtext{↓} & 39.00\redtext{↓} & 56.07\greentext{↑} & 43.67\redtext{↓} \\
\textbf{Aug+Sel} & 13.16\redtext{↓} & 25.51\greentext{↑} & 22.78\greentext{↑} & 29.65\greentext{↑} & 22.78\greentext{↑} & 10.25\redtext{↓} & 12.47\greentext{↑} & 15.96\redtext{↓} & 16.15\redtext{↓} & 13.71\redtext{↓} & 37.80\greentext{↑} & 47.13\redtext{↓} & 49.33\greentext{↑} & 54.27\redtext{↓} & 47.13\greentext{↑} \\
\textbf{Cle+Sel} & 12.39\redtext{↓} & 21.47\greentext{↑} & 20.67\redtext{↓} & 25.54\redtext{↓} & 20.02\redtext{↓} & 13.20\greentext{↑} & 17.39\greentext{↑} & 18.75\redtext{↓} & 16.70\redtext{↓} & 16.51\greentext{↑} & 37.20\greentext{↑} & 48.07\redtext{↓} & 45.60\greentext{↑} & 54.93\greentext{↑} & 46.45\greentext{↑} \\
\midrule
\multicolumn{16}{l}{\framecolorbox[17.9cm][l]{gray!30}{gray!30}{\textbf{CSS (Code Smells)}}} \\ \midrule
\textbf{Ori} & 5.45 & 5.15 & 5.69 & 5.42 & 5.43 & 7.00 & 5.44 & 6.73 & 5.94 & 6.28 & 2.36 & 0.81 & 1.29 & 2.17 & 1.66 \\
\textbf{Syn+Aug} & 6.53\greentext{↑} & 6.77\greentext{↑} & 6.81\greentext{↑} & 6.74\greentext{↑} & \underline{6.71}\greentext{↑} & 7.10\greentext{↑} & 6.50\greentext{↑} & 7.04\greentext{↑} & 7.36\greentext{↑} & 7.00\greentext{↑} & 3.71\greentext{↑} & 4.25\greentext{↑} & 4.16\greentext{↑} & 3.86\greentext{↑} & \underline{4.00}\greentext{↑} \\
\textbf{Syn+Cle} & 6.22\greentext{↑} & 5.73\greentext{↑} & 5.66\redtext{↓} & 6.22\greentext{↑} & 5.96\greentext{↑} & 6.38\redtext{↓} & 5.72\greentext{↑} & 6.62\redtext{↓} & 6.51\greentext{↑} & 6.31\greentext{↑} & 4.20\greentext{↑} & 3.53\greentext{↑} & 3.02\greentext{↑} & 3.15\greentext{↑} & 3.48\greentext{↑} \\
\textbf{Syn+Sel} & 6.12\greentext{↑} & 5.83\greentext{↑} & 6.47\greentext{↑} & 6.30\greentext{↑} & 6.18\greentext{↑} & 5.83\redtext{↓} & 6.32\greentext{↑} & 6.36\redtext{↓} & 6.16\greentext{↑} & 6.17\redtext{↓} & 3.70\greentext{↑} & 2.76\greentext{↑} & 3.23\greentext{↑} & 1.97\redtext{↓} & 2.92\greentext{↑} \\
\textbf{Syn+Ref} & 6.71\greentext{↑} & 6.54\greentext{↑} & 6.68\greentext{↑} & 6.94\greentext{↑} & \textbf{6.72}\greentext{↑} & 7.63\greentext{↑} & 7.15\greentext{↑} & 8.02\greentext{↑} & 6.97\greentext{↑} & \textbf{7.44}\greentext{↑} & 4.01\greentext{↑} & 3.88\greentext{↑} & 4.40\greentext{↑} & 4.13\greentext{↑} & \textbf{4.11}\greentext{↑} \\
\textbf{Ref+Aug} & 5.73\greentext{↑} & 5.96\greentext{↑} & 6.60\greentext{↑} & 5.71\greentext{↑} & 6.00\greentext{↑} & 6.10\redtext{↓} & 6.94\greentext{↑} & 7.32\greentext{↑} & 7.01\greentext{↑} & 6.84\greentext{↑} & 2.70\greentext{↑} & 2.86\greentext{↑} & 2.88\greentext{↑} & 2.66\greentext{↑} & 2.78\greentext{↑} \\
\textbf{Ref+Cle} & 6.04\greentext{↑} & 6.11\greentext{↑} & 6.50\greentext{↑} & 6.15\greentext{↑} & 6.20\greentext{↑} & 6.72\redtext{↓} & 6.65\greentext{↑} & 7.58\greentext{↑} & 6.75\greentext{↑} & 6.93\greentext{↑} & 2.54\greentext{↑} & 1.71\greentext{↑} & 2.22\greentext{↑} & 2.14\redtext{↓} & 2.15\greentext{↑} \\
\textbf{Ref+Sel} & 5.65\greentext{↑} & 5.88\greentext{↑} & 5.66\redtext{↓} & 5.73\greentext{↑} & 5.73\greentext{↑} & 7.39\greentext{↑} & 7.53\greentext{↑} & 7.20\greentext{↑} & 7.03\greentext{↑} & \underline{7.29}\greentext{↑} & 3.70\greentext{↑} & 1.78\greentext{↑} & 3.24\greentext{↑} & 1.88\redtext{↓} & 2.65\greentext{↑} \\
\textbf{Aug+Cle} & 5.83\greentext{↑} & 5.79\greentext{↑} & 6.20\greentext{↑} & 5.95\greentext{↑} & 5.94\greentext{↑} & 7.25\greentext{↑} & 7.16\greentext{↑} & 7.36\greentext{↑} & 6.42\greentext{↑} & 7.05\greentext{↑} & 2.74\greentext{↑} & 2.38\greentext{↑} & 3.59\greentext{↑} & 2.27\greentext{↑} & 2.75\greentext{↑} \\
\textbf{Aug+Sel} & 5.68\greentext{↑} & 5.41\greentext{↑} & 5.90\greentext{↑} & 5.91\greentext{↑} & 5.73\greentext{↑} & 7.11\greentext{↑} & 7.03\greentext{↑} & 6.98\greentext{↑} & 7.09\greentext{↑} & 7.05\greentext{↑} & 3.62\greentext{↑} & 1.81\greentext{↑} & 3.22\greentext{↑} & 1.90\redtext{↓} & 2.64\greentext{↑} \\
\textbf{Cle+Sel} & 5.05\redtext{↓} & 5.27\greentext{↑} & 5.61\redtext{↓} & 5.16\redtext{↓} & 5.27\redtext{↓} & 6.76\redtext{↓} & 7.31\greentext{↑} & 6.44\redtext{↓} & 7.00\greentext{↑} & 6.88\greentext{↑} & 3.70\greentext{↑} & 1.97\greentext{↑} & 2.34\greentext{↑} & 1.83\redtext{↓} & 2.46\greentext{↑} \\
\midrule
\multicolumn{16}{l}{\framecolorbox[17.9cm][l]{gray!30}{gray!30}{\textbf{MI (Code Maintainability)}}} \\ \midrule
\textbf{Ori} & 58.87 & 56.45 & 59.42 & 59.00 & 58.44 & 57.73 & 53.37 & 55.73 & 58.79 & 56.41 & 76.71 & 79.17 & 75.74 & 88.20 & 79.96 \\
\textbf{Syn+Aug} & 52.78\redtext{↓} & 54.23\redtext{↓} & 55.15\redtext{↓} & 53.99\redtext{↓} & 54.04\redtext{↓} & 49.54\redtext{↓} & 45.64\redtext{↓} & 51.25\redtext{↓} & 51.55\redtext{↓} & 49.50\redtext{↓} & 77.54\greentext{↑} & 77.87\redtext{↓} & 80.80\greentext{↑} & 75.37\redtext{↓} & 77.90\redtext{↓} \\
\textbf{Syn+Cle} & 53.14\redtext{↓} & 54.62\redtext{↓} & 55.15\redtext{↓} & 55.21\redtext{↓} & 54.53\redtext{↓} & 53.52\redtext{↓} & 51.85\redtext{↓} & 53.05\redtext{↓} & 53.10\redtext{↓} & 52.98\redtext{↓} & 79.25\greentext{↑} & 78.22\redtext{↓} & 78.61\greentext{↑} & 80.27\redtext{↓} & 79.09\redtext{↓} \\\
\textbf{Syn+Sel} & 55.54\redtext{↓} & 58.58\greentext{↑} & 56.89\redtext{↓} & 57.01\redtext{↓} & 57.01\redtext{↓} & 54.81\redtext{↓} & 52.23\redtext{↓} & 54.71\redtext{↓} & 51.32\redtext{↓} & 53.27\redtext{↓} & 74.17\redtext{↓} & 83.01\greentext{↑} & 73.11\redtext{↓} & 82.13\redtext{↓} & 78.11\redtext{↓} \\
\textbf{Syn+Ref} & 53.26\redtext{↓} & 53.26\redtext{↓} & 54.96\redtext{↓} & 55.00\redtext{↓} & 54.12\redtext{↓} & 52.40\redtext{↓} & 50.07\redtext{↓} & 54.56\redtext{↓} & 51.49\redtext{↓} & 52.13\redtext{↓} & 80.13\greentext{↑} & 77.62\redtext{↓} & 79.99\greentext{↑} & 81.49\redtext{↓} & 79.81\redtext{↓} \\
\textbf{Ref+Aug} & 53.70\redtext{↓} & 54.87\redtext{↓} & 56.81\redtext{↓} & 54.75\redtext{↓} & 55.03\redtext{↓} & 52.15\redtext{↓} & 55.07\greentext{↑} & 55.85\greentext{↑} & 52.77\redtext{↓} & 53.96\redtext{↓} & 71.78\redtext{↓} & 75.71\redtext{↓} & 75.77\greentext{↑} & 76.98\redtext{↓} & 75.06\redtext{↓} \\
\textbf{Ref+Cle} & 61.85\greentext{↑} & 62.08\greentext{↑} & 60.12\greentext{↑} & 61.89\greentext{↑} & 61.49\greentext{↑} & 62.36\greentext{↑} & 57.25\greentext{↑} & 56.77\greentext{↑} & 60.51\greentext{↑} & \underline{59.22}\greentext{↑} & 85.17\greentext{↑} & 86.28\greentext{↑} & 82.89\greentext{↑} & 88.11\redtext{↓} & 85.61\greentext{↑} \\
\textbf{Ref+Sel} & 60.77\greentext{↑} & 60.01\greentext{↑} & 67.45\greentext{↑} & 65.40\greentext{↑} & \underline{63.41}\greentext{↑} & 59.94\greentext{↑} & 59.49\greentext{↑} & 62.24\greentext{↑} & 59.73\greentext{↑} & \textbf{60.35}\greentext{↑} & 89.17\greentext{↑} & 91.44\greentext{↑} & 88.05\greentext{↑} & 92.90\greentext{↑} & \textbf{90.39}\greentext{↑} \\
\textbf{Aug+Cle} & 58.57\redtext{↓} & 57.77\greentext{↑} & 58.72\redtext{↓} & 61.76\greentext{↑} & 59.21\greentext{↑} & 57.28\redtext{↓} & 55.42\greentext{↑} & 56.20\greentext{↑} & 53.00\redtext{↓} & 55.48\redtext{↓} & 74.09\redtext{↓} & 76.56\redtext{↓} & 70.31\redtext{↓} & 78.14\redtext{↓} & 74.78\redtext{↓} \\
\textbf{Aug+Sel} & 58.25\redtext{↓} & 60.63\greentext{↑} & 57.03\redtext{↓} & 61.01\greentext{↑} & 59.23\greentext{↑} & 57.93\greentext{↑} & 53.06\redtext{↓} & 57.55\greentext{↑} & 55.77\redtext{↓} & 56.08\redtext{↓} & 84.15\greentext{↑} & 90.60\greentext{↑} & 83.85\greentext{↑} & 92.07\greentext{↑} & \underline{87.67}\greentext{↑} \\
\textbf{Cle+Sel} & 69.38\greentext{↑} & 64.73\greentext{↑} & 70.38\greentext{↑} & 62.82\greentext{↑} & \textbf{66.83}\greentext{↑} & 58.29\greentext{↑} & 59.83\greentext{↑} & 54.50\redtext{↓} & 56.21\redtext{↓} & 57.21\greentext{↑} & 84.17\greentext{↑} & 82.83\greentext{↑} & 81.08\greentext{↑} & 92.50\greentext{↑} & 85.15\greentext{↑} \\
\midrule
\bottomrule
\end{tabular}
\end{adjustbox}
\begin{tablenotes}
    \scriptsize
    \item[*] \parbox{0.75\linewidth}{
    The combination of data synthesis and data augmentation is denoted as Syn+Aug and analogous notation is used for other technique combinations.
    }
\end{tablenotes}
\end{threeparttable}
\end{table*}

\textbf{Approach.}
In this RQ, we investigate the combined effects of different training data optimization techniques on LLM-based code generation. 
To this end, we systematically construct combined training datasets by sequentially applying two data optimization techniques, followed by fine-tuning the model on the resulting dataset. 
Specifically, we evaluate all 10 pairwise combinations ($C_5^2$), while keeping all other training and evaluation settings identical to those used in RQ1. 
In total, this experimental setup yields 120 model configurations (10 combinations $\times$ 3 benchmarks $\times$ 4 models).
The detailed combination strategy is as follows:

\textit{(1) Combination Scope (Number of Techniques).} 
We focus on pairwise combinations, as they represent the fundamental building blocks of more complex multi-technique pipelines. 
Pairwise settings allow us to explicitly analyze whether two techniques exhibit complementary or conflicting effects, while avoiding the exponential growth in the search space and experimental cost associated with higher-order combinations.

\textit{(2) Combination Strategy (Application Method).}
We adopt a sequential application strategy prior to fine-tuning, in which one technique is applied to the training data, followed by another. 
This approach preserves the individual characteristics of each technique and enables the second technique to potentially mitigate defects introduced by the first. 
In contrast, directly merging independently optimized datasets may introduce duplicate samples and artificially inflate the dataset size, while iterative fine-tuning can substantially increase training costs and the risk of knowledge forgetting.

\textit{(3) Combination Order (Priority Design).}
To ensure effective interaction between techniques, we apply them in a fixed priority order: Data Synthesis $\rightarrow$ Data Refactoring $\rightarrow$ Data Augmentation $\rightarrow$ Data Cleaning $\rightarrow$ Data Selection. 
This is because data synthesis and data refactoring directly modify code and therefore operate at the earliest stage; among them, data synthesis has the highest priority as it relies solely on natural-language problem descriptions rather than existing code. 
Among the remaining techniques, data augmentation may introduce noise, which can subsequently be mitigated by data cleaning, while data selection performs best when applied to already cleaned data. 
This priority design ensures that each technique can maximally leverage the outputs of preceding steps, leading to more coherent and effective combined training datasets.




\smallskip
\noindent
\textbf{Results.} Table~\ref{tab:RQ2} presents the metrics of ten combination techniques across twelve experimental settings. 
The “Avg.” column shows the average performance over four models, with bold indicating the best-performing technique and underlined indicating the second-best. 
"\greentext{↑}" and "\redtext{↓}" in the table indicate results higher or lower than those fine-tuned on the original dataset, respectively. 
For brevity, for example, we denote the combination of data synthesis and data refactoring as Syn+Ref, and apply the same notation to other technique combinations.

\textbf{Functional Correctness.}
The results show that the pairwise combinations fluctuate generally and do not lead to further improvements in functional correctness.
In most cases, the effectiveness of a combination is close to that of the better individual technique in this pair, and sometimes even slightly lower. 
Combinations involving data synthesis maintain relatively high correctness, achieving 30.48\%–42.66\% \textit{Pass@1} and 15.35\%–18.92\% \textit{AvgPassRatio} relative improvements, but their effectiveness rarely exceeds that of using data synthesis alone. 
Among these combinations, the combination of data synthesis and data refactoring performs the best, achieving a higher average \textit{Pass@1} than using data synthesis alone on the APPS and MBPP benchmarks, with relative improvements of 62.50\% and 12.58\%, respectively. 
On the MBPP benchmark, its average \textit{AvgPassRatio} also surpasses that of using data synthesis alone, with relative improvements of 11.07\%.

This can be explained by the intrinsic characteristics of synthesized data, which comprise high-quality, well-aligned samples that strongly reinforce correct solution patterns. 
When such data are subsequently subjected to data cleaning or data augmentation, these operations may inadvertently disrupt logical structures or stylistic consistency, or introduce irrelevant noise, thereby weakening the correctness signal. 
In contrast, data refactoring enforces code standardization and structural clarity without altering the underlying semantics, making it particularly complementary to data synthesis.
Combinations involving other techniques (e.g., Ref+Aug and Aug+Sel) do not exhibit stable or substantial improvements in functional correctness, with their effects varying across models and benchmarks. 
This variability likely stems from the relatively mild nature of these techniques, such as readability, oriented refactoring, or simple data selection, which only marginally modify the training data. 
As a result, their combined influence on overall data quality remains limited.
We further investigate the relationship between the extent of training data modification and the effectiveness of technique combinations in Section~\ref{subsec:RQ3}.


\finding{V}{Combining training data optimization techniques does not lead to substantial additional improvements in the functional correctness of generated code. Nevertheless, combinations involving data synthesis consistently maintain relatively high accuracy. Among them, the synthesis–refactoring combination (Syn+Ref) achieves the best performance, with average relative improvements of 36.82\% in \textit{Pass@1} and 18.15\% in \textit{AvgPassRatio}.}

\textbf{Code Smells. }
Nearly all combinations of training data optimization techniques reduce code smells.
With the exception of Cle+Sel on APPS and Syn+Sel on CodeContests, all other combinations improve the CSS by an average of 18.02\%–63.27\%. 
Compared with applying each technique individually, almost all combinations, except Cle+Sel, achieve higher CSS on APPS. 
In CodeContests, however, combinations rarely yield further improvements, with only Syn+Ref outperforming both individual techniques. On MBPP, the results fall between these two extremes, where Syn+Aug, Syn+Ref, Ref+Aug, Ref+Cle, and Aug+Cle outperform their corresponding individual techniques.
Overall, most combinations maintain relatively high CSS, indicating that the combination process generally does not introduce additional code smells and preserves static code quality.

In particular,  Syn+Ref demonstrates superior effectiveness. 
It not only achieves the highest CSS across all benchmarks but also outperforms data synthesis or data refactoring alone in 10 out of 12 scenarios, indicating a strong synergistic effect between the two techniques. 
Data synthesis contributes high-quality code with correct solutions and consistent stylistic patterns, while data refactoring further enforces compliance with coding standards and structural clarity. Together, they enable the generation of code that exhibits both high functional correctness and substantially fewer code smells.
On the other hand, Cle+Sel performs worse than applying either technique individually in 10 out of 12 scenarios and even reduces the CSS in 7 of these cases. One plausible explanation is that consecutive filtering by data cleaning and data selection produces training data that are overly homogeneous in terms of style and structure, thereby reducing structural diversity. 
As a result, the fine-tuned models tend to overuse a limited set of fixed code patterns, which ultimately leads to an increase in code smells.


\finding{VI}{Most combinations help reduce code smells; however, their ability to outperform individual techniques depends on the dataset. 
The synthesis–refactoring combination (Syn+Ref) achieves the strongest effect, delivering the highest average relative improvement of 63.27\% in Code Smell Score compared to the individual technique.}

\textbf{Code Maintainability. }
Pairwise combinations involving data refactoring, data cleaning, and data selection demonstrate significant advantages in enhancing code maintainability.
Among them, the Ref+Sel outperforms both individual techniques across all three benchmarks, with average MI relative improvements of 8.50\%, 6.98\%, and 13.04\%. The Ref+Cle also shows further improvement in MBPP with an average MI relative improvement of 7.07\%. The Cle+Sel surpasses the individual technique results in APPS and MBPP, with average MI relative improvements of 14.36\% and 6.49\%. 
These combinations typically generate code with higher code maintainability than that produced by any single technique, because all three techniques aim to optimize or filter the quality of the code in the dataset. Their combination further enhances code quality from complementary perspectives, leading to superior results compared with using each technique individually.

Notably, although the Cle+Sel performs poorly in terms of CSS, it still produces code with relatively high code maintainability, revealing a trade-off between code smells and code maintainability. 
When dataset diversity is excessively reduced and code structures become overly simplified, the learning space available to the model is compressed. Consequently, the model may generate code with lower complexity and higher code maintainability, while simultaneously exhibiting more code smells due to repetitive or rigid structural patterns. 
This observation suggests that, when designing training data optimization pipelines, it is crucial to balance dataset diversity and code complexity to avoid degrading certain aspects of code quality while improving others.

In contrast, the negative impact of data synthesis and data augmentation on code maintainability continues to appear in their combinations. Although these techniques increase diversity and data scale, their generated codes often lack structural regularity and conciseness, likely introducing redundancy and non-standard constructs. 
Even when combined with data cleaning, data refactoring, or data selection, these structural weaknesses are difficult to eliminate. 

\finding{VII}{Pairwise combinations involving data refactoring, data cleaning, and data selection significantly improve code maintainability by average relative improvements of 5.76\%--9.51\%. 
The negative influence of data augmentation and data synthesis persists when these techniques are included in combinations.}

\begin{table*}[h]
\centering
\footnotesize
\caption{Model sensitivity to the combined training data optimization}
\label{tab:model_comparison_combination}
\setlength{\tabcolsep}{4pt}
\begin{adjustbox}{width=\textwidth}
\begin{tabular}{l cc c cc c cc} 
\toprule
\midrule
\multirow{2}{*}{\textbf{Technique}} 
& \multicolumn{2}{c}{\textbf{Pass@1}} 
& \multicolumn{2}{c}{\textbf{AvgPassRatio}} 
& \multicolumn{2}{c}{\textbf{CSS}} 
& \multicolumn{2}{c}{\textbf{MI}} \\
\cmidrule(lr){2-3} \cmidrule(lr){4-5} \cmidrule(lr){6-7} \cmidrule(lr){8-9}
 & QW Avg. & LM Avg. & QW Avg. & LM Avg. & QW Avg. & LM Avg. & QW Avg. & LM Avg. \\
\midrule
\textbf{Syn+Aug} & \textbf{+5.99} & +2.28 & \textbf{+6.61} & +1.84 & \textbf{+1.76} & +1.14 & \textbf{-6.06} & -2.86 \\
\textbf{Syn+Cle} & \textbf{+5.80} & +1.90 & \textbf{+5.85} & +2.32 & \textbf{+0.99} & +0.60 & \textbf{-3.62} & -1.91 \\
\textbf{Syn+Sel} & \textbf{+4.08} & +2.42 & \textbf{+3.87} & +3.23 & \textbf{+0.74} & +0.53 & -1.78 & \textbf{-2.50} \\
\textbf{Syn+Ref} & \textbf{+4.92} & +3.52 & \textbf{+5.01} & +3.73 & \textbf{+1.78} & +1.49 & \textbf{-4.34} & -1.48 \\
\textbf{Ref+Aug} & \textbf{+0.98} & -0.57 & \textbf{+0.17} & -0.97 & \textbf{+1.04} & +0.47 & \textbf{-4.14} & -3.02 \\
\textbf{Ref+Cle} & \textbf{+1.72} & +0.92 & \textbf{+2.09} & +1.31 & \textbf{+0.76} & +0.51 & +3.52 & \textbf{+4.16} \\
\textbf{Ref+Sel} & +1.57 & \textbf{+1.91} & +1.42 & \textbf{+1.86} & \textbf{+0.82} & +0.72 & +5.67 & \textbf{+7.24} \\
\textbf{Aug+Cle} & \textbf{+0.20} & -0.05 & \textbf{+0.48} & +0.01 & \textbf{+0.84} & +0.74 & \textbf{-2.06} & -1.51 \\
\textbf{Aug+Sel} & -0.40 & \textbf{+1.51} & +0.68 & \textbf{+1.43} & \textbf{+0.70} & +0.67 & \textbf{+3.03} & +2.43 \\
\textbf{Cle+Sel} & +0.28 & \textbf{+1.24} & +0.50 & \textbf{+1.19} & \textbf{+0.60} & +0.23 & +3.99 & \textbf{+5.60} \\
\midrule
\bottomrule
\end{tabular}
\end{adjustbox}
\end{table*}

\textbf{Model Sensitivity to Combined Training Data Optimization}. Table~\ref{tab:model_comparison_combination} summarizes the average performance changes of the two model families, where bold values indicate the model family exhibiting larger absolute performance increases or decreases, or the one showing an increase when the two families exhibit divergent trends. 
These results are consistent with Finding IV in the context of combined data optimization techniques. Across 25 out of 36 combination scenarios, Qwen2.5-Coder exhibits larger performance variations than Llama-3.2, indicating higher sensitivity to changes introduced by technique combinations. 
Moreover, in 4 scenarios where the two model families exhibit divergent trends, Qwen2.5-Coder improves in 3 cases while Llama-3.2 experiences performance degradation. 
This suggests that Qwen2.5-Coder is better able to benefit from the complementary effects of combined data optimization techniques.
This behavior can be attributed to the code-specialized nature of Qwen2.5-Coder. 
Having learned a richer and more diverse set of code patterns during pretraining, Qwen2.5-Coder can more effectively exploit the structural and semantic signals introduced by combined data transformations. Consequently, when technique combinations introduce both beneficial modifications and potential noise, Qwen2.5-Coder is more capable of filtering out harmful effects and leveraging useful information, resulting in more stable or improved performance under combined data optimization settings.

\finding{VIII}{Under combined data optimization settings, code-specialized models (Qwen2.5-Coder) are more sensitive yet more resilient than general-purpose models (Llama-3.2): they exhibit larger performance fluctuations but are more likely to benefit from technique combinations, indicating stronger capability to exploit complementary data transformations while filtering out introduced noise.}

\subsection{RQ3: Fine-Grained Correlation Analysis}
\label{subsec:RQ3}

\textbf{Approach.}
To facilitate a deeper interpretation of the results from RQ1 and RQ2, we perform a fine-grained correlation analysis of code generation performance from two perspectives: 
(1) the correlation between the code modification magnitude introduced by a technique and its effectiveness when used individually (corresponding to \textit{RQ1}), as well as its role in the combination setting (corresponding to \textit{RQ2}), and (2) the correlation between the complementarity of individual techniques and their combination effectiveness (corresponding to \textit{RQ2}).
In this research question, we focus on functional correctness metrics for two primary reasons. 
First, functional correctness is the most fundamental and widely adopted evaluation criterion in code generation tasks. 
Second, unlike code quality metrics, functional correctness cannot be readily inferred or explained solely based on a technique's workflow, making empirical evaluation essential.



\textbf{Analysis I: Effect of Code Modification Magnitude.} 
Specifically, we examine the relationship between the code modification magnitude and effectiveness, considering both individual training data optimization techniques and the biases introduced in their combinations.
To quantify the code modification magnitude (i.e., the extent to which an optimized dataset differs from the original dataset), we adopt the \textit{Fréchet Inception Distance (FID)}~\cite{heusel2017gans} as a measure of overall distributional divergence between two datasets.
Although originally proposed for the field of computer vision~\cite{desai2024rsuigm, qin2023adversarial}, FID compares Gaussian approximations of feature distributions and has been successfully extended to other domains, including \textit{Fréchet Audio Distance} for audio and speech processing~\cite{kilgour2018fr}, \textit{Fréchet BERT Distance} for natural language processing~\cite{xiang2021assessing}, and \textit{Fréchet CLIP Distance} for multimedia retrieval~\cite{bithel2023evaluating}.
Following this paradigm, we compute FID with the pre-trained code embeddings model (CodeBERT~\cite{feng2020codebert}) to measure code differences in style, structure, and semantics. 
Formally, given code embeddings of the original dataset $\mathbf{X}$ and the optimized dataset $\mathbf{Y}$, with mean vectors $\mu_X, \Sigma_X$ and $\mu_Y, \Sigma_Y$, respectively, the FID value is computed as:
\begin{equation}
    FID(\mathbf{X}, \mathbf{Y}) = \|\mu_X - \mu_Y\|^2 + \mathrm{Tr}\Big(\Sigma_X + \Sigma_Y - 2(\Sigma_X \Sigma_Y)^{1/2}\Big)
    \label{eq:fid}
\end{equation}
where $\|\cdot\|$ denotes the Euclidean norm and $\mathrm{Tr}(\cdot)$ is the trace operator.
A higher FID value indicates greater code modification magnitude.

Corresponding to RQ1, we compute the FID values and identify uniquely solved problem sets (visualized using Venn diagrams) for the five training data optimization techniques across three benchmarks. 
Corresponding to RQ2, we investigate the impact of FID on the combined optimization techniques from both dataset-level and instance-level perspectives. 
At the dataset level, we compute the intersection ratio between the solved sets of a combined technique and the solved sets of its two constituent techniques (referred to as ``Intersection'').
We analyze it jointly with FID to examine how FID influences the effectiveness of combination strategies.
At the instance level, we compute the cosine similarity (based on CodeBERT embeddings) between code generated by a combined technique and that generated by each of its constituent techniques (referred to as ``Similarity'').
We analyze it together with FID to assess how FID affects the structural and stylistic characteristics of code generated by combined techniques.

\textbf{Analysis II: Effect of Technique Complementarity.} 
Corresponding to RQ2, we examine technique complementarity by computing the size of the solved problem set produced by each combined optimization technique (referred to as ``Combined Set'') and comparing it with the size of the union of the solved sets produced by its two constituent techniques (referred to as ``Union Set'').
For each benchmark, we calculate the average sizes of the combined sets and their corresponding union sets, and then compute the Pearson correlation coefficient~\cite{benesty2009pearson} along with the associated p-value to quantify the statistical relationship between them.
In addition, using the performance of QW-3B on the APPS benchmark as an illustrative case, we visualize the overlaps between combined sets and union sets using Venn diagrams. 
These visualizations provide an intuitive understanding of the extent to which the effectiveness of combined techniques aligns with the aggregated effectiveness of their constituent techniques.

\begin{figure*}[]
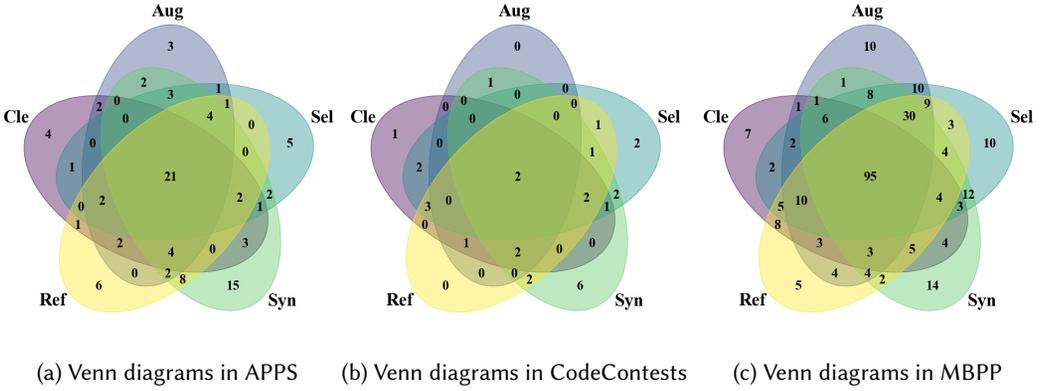

\centering
\begin{minipage}{0.33\textwidth}
    \centering
    \includegraphics[width=\linewidth]{figures/APPS.pdf}
    \subcaption{Venn diagrams in APPS}
    \label{fig:venn_individual:apps}
\end{minipage}%
\hfill%
\begin{minipage}{0.33\textwidth}
    \centering
    \includegraphics[width=\linewidth]{figures/Codecontests.pdf}
    \subcaption{Venn diagrams in CodeContests}
    \label{fig:venn_individual:codecontests}
\end{minipage}
\hfill%
\begin{minipage}{0.33\textwidth}
    \centering
    \includegraphics[width=\linewidth]{figures/MBPP.pdf}
    \subcaption{Venn diagrams in MBPP}
    \label{fig:venn_individual:mbpp}
\end{minipage}
\caption{Uniquely solved instances across five individual training data optimization techniques}
\label{fig:venn_individual}
\end{figure*}

\begin{figure*}[]
\centering
\makebox[\textwidth]{%
  \includegraphics[width=\textwidth,height=\textheight,keepaspectratio]{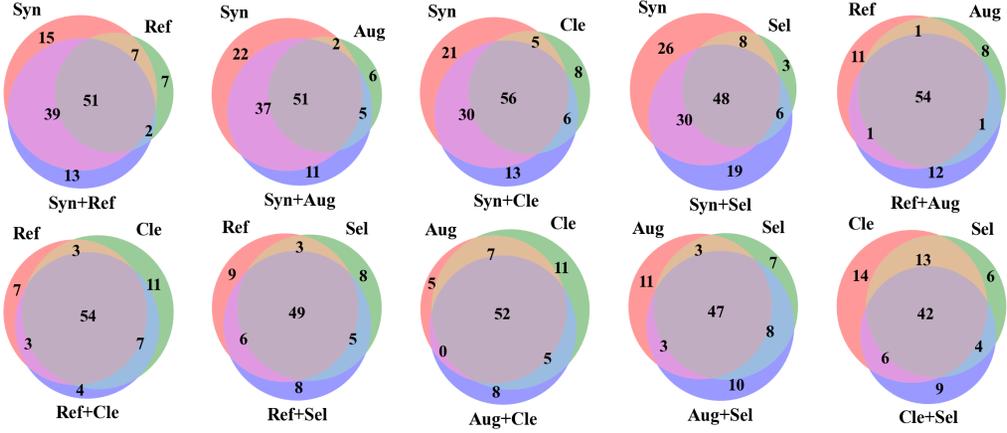}%
}
\caption{Uniquely solved instances for combined techniques and their constituent techniques}
\label{fig:venn_combine}
\end{figure*}

\smallskip
\noindent
\textbf{Results.} 
Tables~\ref{tab:modification}, \ref{tab:similarity}, and \ref{tab:correlation}, together with Figures~\ref{fig:venn_individual} and \ref{fig:venn_combine}, summarize the results of \textbf{Analysis I} and \textbf{Analysis II}. 
Table~\ref{tab:modification} presents the FID values, which quantify the code modification magnitudes, while Figure~\ref{fig:venn_individual} illustrates the uniquely solved sets for individual techniques on LM-3B across three benchmarks.
Table~\ref{tab:similarity} further presents the ``Intersection'' and ``Similarity'' metrics between combined techniques and their constituent techniques.
The ``Difference'' column reports values computed as the first technique minus the second technique.
Positive (negative) values indicate a bias toward the first (second) technique, and differences with an absolute value greater than 0.1 are highlighted in bold. 
Table~\ref{tab:correlation} reports the Pearson correlation coefficients between the average sizes of combined sets and union sets across three benchmarks.
In addition, Figure~\ref{fig:venn_combine} visualizes these relationships using Venn diagrams for the ten combined techniques on APPS with QW-3B.

\begin{table*}[t]
    \centering
    \tabcolsep=8.0mm
    \normalsize
    \captionof{table}{Code modification magnitude of five individual training data optimization techniques in FID metric}
    \label{tab:modification}
    \vspace{0.2em}
    \begin{tabular}{lccc}
        \toprule
        \midrule
        \textbf{Technique} & \textbf{APPS} & \textbf{CodeContests} & \textbf{MBPP} \\
        \midrule
        \textbf{Data Augmentation} & 0.6997 & 1.0868 & 2.2728 \\
        \textbf{Data Cleaning} & 0.3901 & 0.0331 & 0.0139 \\
        \textbf{Data Selection} & \underline{1.4041} & \underline{1.8532} & \underline{3.1822} \\
        \textbf{Data Synthesis} & \textbf{2.5062} & \textbf{2.1012} & \textbf{8.3825} \\
        \textbf{Data Refactoring} & 1.1347 & 0.4310 & 3.0933 \\
        \midrule
        \bottomrule
    \end{tabular}
    \vspace{0.3em}
\end{table*}

\textbf{Effect of Code Modification Magnitude in Individual Techniques.}
Table~\ref{tab:modification} shows that data synthesis introduces the largest code modification across all three benchmarks, while data selection and data refactoring also result in substantial changes. 
This observation is consistent with the results reported in Table~\ref{tab:RQ1} of RQ1, indicating that the improvement in functional correctness brought by training data optimization techniques is closely related to the effective magnitude of dataset modification. 
Figure~\ref{fig:venn_individual} also illustrates this finding, where data synthesis relatively solves more unique instances (i.e., 15, 6, and 14 for LM-3B in APPS, CodeContests, and MBPP, respectively) than other techniques. 
In contrast, data cleaning and data augmentation introduce relatively minor code modification magnitude to the original dataset. 
Data cleaning may remove only a small amount of low-quality code, while the transformation rules used in data augmentation may be overly simple, resulting in limited overall dataset improvement and only marginal impacts on improving model effectiveness. 
These findings suggest that, when designing training data optimization techniques, overly minor modifications should be avoided in favor of approaches that introduce broader and more substantial changes for low-quality datasets.

\finding{IX}{Techniques that introduce larger effective modifications to the low-quality training dataset tend to yield greater improvements in LLM-based code generation effectiveness. 
This finding reinforces \textbf{Finding I}, which shows that data synthesis performs best among all individual techniques.}

\begin{table*}[t]
    \centering
    \tabcolsep=2.5mm
    \footnotesize
    \caption{The impact of code modification magnitude on combined techniques in ``Intersection'' and ``Similarity''}
    \label{tab:similarity}
    \begin{adjustbox}{width=\textwidth}
    \begin{tabular}{l cc cc cc} 
        \toprule
        \midrule
        \textbf{Combination}
        & \multicolumn{2}{c}{\textbf{Tech$_1$}} 
        & \multicolumn{2}{c}{\textbf{Tech$_2$}} 
        & \multicolumn{2}{c}{\textbf{Difference}}  \\
        \cmidrule(lr){2-3} \cmidrule(lr){4-5} \cmidrule(lr){6-7}
        \textbf{(Tech$_1$ + Tech$_2$)}
        & Intersection & Similarity & Intersection & Similarity & Intersection & Similarity \\
        \midrule
        \textbf{Syn+Ref} & 0.7953 & 0.6867 & 0.5194 & 0.4307 & \textbf{0.2759} & \textbf{0.2560} \\
        \textbf{Syn+Aug} & 0.7560 & 0.5690 & 0.4783 & 0.4099 & \textbf{0.2777} & \textbf{0.1591} \\
        \textbf{Syn+Cle} & 0.8195 & 0.7206 & 0.5435 & 0.4257 & \textbf{0.2760} & \textbf{0.2949} \\
        \textbf{Syn+Sel} & 0.7265 & 0.5268 & 0.6718 & 0.5565 & 0.0547 & -0.0297 \\
        \textbf{Ref+Aug} & 0.7455 & 0.6148 & 0.6716 & 0.6138 & 0.0739 & 0.0010 \\
        \textbf{Ref+Cle} & 0.7220 & 0.6780 & 0.6625 & 0.6989 & 0.0595 & -0.0209 \\
        \textbf{Ref+Sel} & 0.6669 & 0.5229 & 0.7841 & 0.7450 & \textbf{-0.1172} & \textbf{-0.2221} \\
        \textbf{Aug+Cle} & 0.6552 & 0.6566 & 0.6953 & 0.6066 & -0.0401 & 0.0500 \\
        \textbf{Aug+Sel} & 0.6006 & 0.4733 & 0.7666 & 0.6985 & \textbf{-0.1660} & \textbf{-0.2252} \\
        \textbf{Cle+Sel} & 0.6430 & 0.5914 & 0.7754 & 0.6940 & \textbf{-0.1324} & \textbf{-0.1026} \\
        \midrule
        \bottomrule
    \end{tabular}
    \end{adjustbox}
\end{table*}

\textbf{Effect of Code Modification Magnitude in Combined Techniques.}
Table~\ref{tab:similarity} shows that, according to both the ``Intersection'' and ``Similarity'' metrics, most combined techniques exhibit a pronounced tendency toward data synthesis or data selection. 
Notably, these two techniques also introduce the largest code modification magnitude to the original training data (as shown in Table~\ref{tab:modification}). 
This observation affirms that techniques that more extensively modify the original low-quality dataset tend to dominate both the combination effectiveness and the structural and stylistic characteristics of the generated code. 
This effect can be explained by the fact that, when two techniques are combined, large-scale code modifications are more prominently reflected in the resulting optimized training data, causing the generated code to inherit and amplify the properties of the dominant technique. 
These findings further imply that, when combining techniques from different categories, the tendency of the combined effect can be modulated by controlling the extent to which each technique modifies the training data, thereby enabling more fine-grained trade-offs among different code quality metrics.

\finding{X}{Techniques that introduce larger effective code modifications to the original training dataset tend to play a dominant role when used in combination. 
This finding reinforces \textbf{Finding V}, which indicates that combinations involving data synthesis achieve larger gains in LLM-based code generation effectiveness.}

\begin{table*}[b]
    \centering
    \tabcolsep=8.5mm
    \captionof{table}{Pearson correlation between the sizes of union set and combined set}
    \label{tab:correlation}
    \vspace{0.2em}
    \begin{tabular}{lcc}
        \toprule
        \midrule
        \textbf{Benchmark} & \textbf{Correlation Coefficient} & \textbf{p-value} \\
        \midrule
        \textbf{APPS}          & 0.9769 & 0.00000\textsuperscript{***} \\
        \textbf{CodeContests}  & 0.6897 & 0.00000\textsuperscript{***} \\
        \textbf{MBPP}          & 0.9667 & 0.00000\textsuperscript{***} \\
        \midrule
        \bottomrule
    \end{tabular}
    \vspace{0.3em}
\end{table*}

\textbf{Effect of Technique Complementarity}. 
Table~\ref{tab:correlation} shows that, across all three benchmarks, the size of the combined set (the pairwise combined optimization technique) is positively correlated with the size of the union set (the constituent techniques) with high statistical significance (\textit{p-value} < 0.00001).
This indicates that the effectiveness of a combination in functional correctness largely depends on the union set of its constituent techniques. 
The Venn diagrams in Figure~\ref{fig:venn_combine} also confirm this finding. 
When data synthesis is combined with any of the other four techniques, the union sets are larger on average, reaching 122.75, and the number of unique correct cases from each technique is also larger, resulting in greater improvements in functional correctness metrics as shown in Table~\ref{tab:RQ2}. 
In contrast, combined techniques that do not involve data synthesis yield smaller union sets on average (80.83) and exhibit greater overlap between constituent techniques, which results in relatively modest gains in functional correctness. 
Therefore, when designing combinations of training data optimization techniques, it is advisable to first evaluate individual techniques and then prioritize combining techniques with larger union sets to ensure the combined effectiveness.
Furthermore, we observe that the union sets remain substantially larger than the combined sets, indicating that current combinations do not fully exploit the effectiveness of individual techniques. 
This limitation is discussed in more detail in Section~\ref{subsec:upper_limit}.

\finding{XI}{The functional correctness achieved by a combined technique is positively correlated with the union set of its constituent techniques, and combinations of dissimilar techniques tend to yield greater effectiveness than those of similar techniques. 
This observation further reinforces \textbf{Finding V}, which shows that the combination of data synthesis and data refactoring (i.e., Syn+Ref) achieves the best performance.}

\section{Discussion}
\label{sec:dis}
\subsection{Case Study: Effect of Data Optimization on Generated Code}
\begin{figure*}[htbp]
\centering

\begin{subfigure}[t]{\textwidth}
    \centering
    \includegraphics[width=\textwidth,height=\textheight,keepaspectratio]{figures/correct.pdf}
    \caption{Cases where the model generates correct code (i.e., ``Ori code'') when trained on the original dataset}
    \label{fig:casestudy_correct}
\end{subfigure}

\vspace{1em}

\begin{subfigure}[t]{\textwidth}
    \centering
    \includegraphics[width=\textwidth,height=\textheight,keepaspectratio]{figures/wrong.pdf}
    \caption{Cases where the model generates incorrect code (i.e., ``Ori code'') when trained on the original dataset}
    \label{fig:casestudy_wrong}
\end{subfigure}

\caption{Case study of code generated by QW-3B trained on the original training dataset and five optimized training datasets on APPS}
\label{fig:casestudy}
\end{figure*}

To gain a deeper understanding of how these techniques affect code generation, we qualitatively analyze two groups of generation examples generated by the QW-3B on APPS. 
Each group includes code generated by models trained on the original training data and five optimized training datasets. 
The corresponding examples are presented in Figure~\ref{fig:casestudy}.

Across both groups, data synthesis exhibits the most pronounced and consistent impact on the style and structure of the generated code. 
In terms of code style, code produced under data synthesis typically adopts variable names with richer semantic information, such as \textit{home\_guest} and \textit{count} in Figure~\ref{fig:casestudy_correct}, and \textit{steps} in Figure~\ref{fig:casestudy_wrong}, which helps to clarify the semantic roles of variables. 
In terms of code structure, data synthesis consistently yields a highly uniform organization, where all solutions are encapsulated within a \textit{solve} function and invoked at the end of the program. 
This stylistic and structural consistency can be attributed to the high semantic density of the synthesized training data, which enables the model to better understand variable usage and program logic, thereby explaining the strong functional correctness achieved by data synthesis. 
Moreover, the standardized function structure reflects the well-organized nature of the synthesized code, allowing the model to generate code in a more regularized manner, which further accounts for its superior performance in reducing code smells.
Despite these advantages, code generated under data synthesis tends to exhibit longer code length and more complex solution structures. 
In both example groups, data synthesis produces the longest code. 
For instance, the example in Figure~\ref{fig:casestudy_correct} introduces a two-dimensional array, while the example in Figure~\ref{fig:casestudy_wrong} relies on recursive function calls. 
Although such designs enhance expressive power, they also increase code complexity, which partially undermines code maintainability. 
This observation helps explain the relatively weaker performance of data synthesis with respect to maintainability-related metrics.

In contrast, the remaining training data optimization techniques demonstrate different optimization characteristics. 
Data selection effectively reduces code smells by decomposing complex \textit{if} conditions and simplifying \textit{while} loop logic, while maintaining relatively low code length and complexity. 
Data cleaning improves code maintainability primarily by simplifying arithmetic expressions and optimizing \textit{for} loop structures. 
Data augmentation introduces only minor modifications to the training data, resulting in generated code that remains highly similar to the original and therefore exhibits no clear optimization effects. 
Data refactoring likewise produces code that is largely similar to the original, but it still demonstrates improvements in code maintainability through the decomposition of \textit{if} conditional statements and the optimization of \textit{for} loop constructs. 
Overall, these techniques tend to preserve the original solution logic and explore a more constrained solution space, which likely limits their impact on the overall functional correctness of the generated code.

\subsection{Upper Bound of Optimization Combination}
\label{subsec:upper_limit}

\begin{table}[htbp]
    \centering
    \caption{Average sizes of union set and combined set with the combined-to-union size ratio}
    \label{tab:upper_limit}
    \resizebox{\textwidth}{!}{
    \begin{tabular}{l ccc ccc ccc}
        \toprule
        \midrule
        \multirow{2}{*}{\textbf{Model}} & \multicolumn{3}{c}{\textbf{APPS}} & \multicolumn{3}{c}{\textbf{CodeContests}} & \multicolumn{3}{c}{\textbf{MBPP}} \\
        \cmidrule(lr){2-4} \cmidrule(lr){5-7} \cmidrule(lr){8-10} 
        & Union & Combined & Proportion & Union & Combined & Proportion & Union & Combined & Proportion \\
        \midrule
        \textbf{LM-1B} & 35.1 & 24.7 & 70.37\% & 9.1  & 5.6  & 61.54\% & 179.8 & 138.7 & 77.14\% \\
        \textbf{LM-3B} & 68.3 & 53.2 & 77.89\% & 20.6 & 14.3 & 69.42\% & 238.9 & 194.8 & 81.54\% \\
        \textbf{QW-1.5B} & 72.2 & 58.3 & 80.75\% & 13.2 & 9.1  & 68.94\% & 281.9 & 231.0 & 81.94\% \\
        \textbf{QW-3B} & 97.6 & 81.5 & 83.50\% & 15.1 & 11.0 & 72.85\% & 306.6 & 258.5 & 84.31\% \\
        \midrule
        \bottomrule
    \end{tabular}
    }
\end{table}
As described in Section~\ref{subsec:RQ3}, we treat the union of solved sets produced by two constituent techniques as the theoretical upper bound on the functional correctness achievable by their combination. 
Table~\ref{tab:upper_limit} reports the average sizes of the union sets and the corresponding combined sets for the ten technique combinations across four models and three benchmarks, together with the ratio of the combined-set size to the union-set size. 
This ratio indicates how closely each combination approaches its theoretical upper bound in terms of functional correctness.

The results show that the current combined techniques still fall significantly short of the theoretical upper bound, with combined sets covering only 60\% to 85\% of the union sets. 
Moreover, the degree to which combined sets approach the upper bound varies with benchmark difficulty, model architecture, and scale. 
Specifically, the higher the benchmark difficulty, the harder it is for combination techniques to reach the upper bound. 
Among the three benchmarks, the easiest MBPP achieves an average ratio of 81.23\%, the most difficult CodeContests achieves 52.96\%, and the medium APPS achieves 78.13\%. 
This indicates that in more challenging benchmarks, current combination techniques struggle to fully leverage the strengths of both constituent optimizations, highlighting the need for more effective combination strategies in complex scenarios.
Regarding model architecture and scale, larger models and code-specific models benefit more from combination techniques. 
Specifically, the 3B-scale models achieve an average ratio of 78.25\%, compared to 73.45\% for the 1B-scale models. 
In addition, code-specific Qwen2.5-Coder models attain 78.72\%, whereas general-purpose Llama-3.2 models reach 61.37\%.
This demonstrates that combination techniques are more effective for larger and code-specific models, underscoring the feasibility of enhancing advanced model performance through data optimization combinations.

\section{Threats to Validity}
\textbf{External Threat}. 
The primary external threat concerns our choices of benchmarks and training data optimization techniques. 
First, our study focuses exclusively on Python programs, as Python is the most widely-used language in code generation research. 
Second, we investigate five training data optimization techniques that are commonly adopted in the code generation domain. 
Moreover, for each optimization category, we select a single representative technique, under the assumption that techniques within the same category follow similar workflows and exhibit comparable characteristics.
In future work, we plan to extend our experimental framework to cover a broader range of training data optimization techniques and to evaluate their effectiveness across multiple programming languages, thereby improving the generalizability of our findings.

\smallskip
\noindent
\textbf{Construct Threat.} 
We identify three construct-related threats. 
First, the choice of evaluation metrics may introduce bias by overlooking certain aspects of code quality, such as readability. 
However, readability is inherently subjective and difficult to measure in a standardized manner. 
Second, the choice of the teacher LLM used in the data synthesis process may influence the observed effectiveness of data synthesis, although we focus on general patterns rather than model-specific behaviors. 
Third, due to computational and cost constraints, we do not repeat training data optimization and fine-tuning multiple times, which may leave residual variance or randomness unaccounted for.

\smallskip
\noindent
\textbf{Internal Threat.} 
The primary internal threat concerns the implementation of the training data optimization techniques. 
To mitigate this risk, we implemented each technique by strictly following the workflows or using the open-source tools provided in the original papers, and the implementations were carefully reviewed by three authors.
In addition, randomness in model fine-tuning and inference may affect the results. 
Although we control this threat by fixing training settings and using low-temperature decoding, residual randomness cannot be completely eliminated.

\section{Conclusion and Future Work}
\label{sec:conclusion}
This paper presents a systematic empirical study that comprehensively analyzes the effectiveness of training data optimization techniques and their combinations for code generation tasks. 
Our results indicate that data synthesis is the most effective technique for improving functional correctness and reducing code smells, but it lags behind data refactoring, data cleaning, and data selection in terms of code maintainability. 
Furthermore, we find that although combined techniques exhibit a clear upper bound in further improving functional correctness, they can substantially enhance overall code quality. 
Among all combinations, data synthesis combined with data refactoring achieves the best overall performance.
Our fine-grained analysis further reveals a strong positive correlation between the sizes of combined sets and their corresponding union sets, and demonstrates that the magnitude of effective code modifications to the training dataset is a key factor influencing the functional correctness of both individual techniques and their combinations.

This work also opens several promising directions for future research: 
(I) The current study focuses on function-level code generation tasks.
Future work could investigate the effectiveness of data optimization techniques in more complex settings, such as multi-function modules or project-level code generation.
(II) Diversifying combination methods. 
Our study primarily examines pairwise combinations of studied training data optimization techniques.
However, the design space of combination strategies remains largely unexplored. 
Future work could consider more advanced approaches, such as multi-technique combinations, adaptive ordering strategies, or post-selection and post-filtering methods, to more fully exploit the potential of training data optimization for code generation.
(III) Validating on larger-scale models. 
The current experiments are conducted on small- to medium-scale models. 
Extending the evaluation to larger-scale LLMs would provide deeper insights into the scalability and generalizability of training data optimization techniques, as well as their interaction with model capacity. 

\begin{acks}
This work was supported by the National Key Research and Development Program of China (Grant No. 2024YFB4506300), the National Natural Science Foundation of China (Grant Nos. 62322208, 12411530122, and 62232001).
\end{acks}

\balance
\bibliographystyle{ACM-Reference-Format}
\bibliography{main}

\end{document}